\documentclass[%
reprint,
superscriptaddress,
preprint numbers,
nofootinbib,
amsmath,amssymb,
aps,
prd
]{revtex4-2}

\usepackage[T1]{fontenc}
\usepackage{graphicx}  
\usepackage{amsmath}   
\usepackage{amssymb}   
\usepackage{braket}
\usepackage{bm}        
\usepackage{float}     
\usepackage{dcolumn}   
\usepackage{xcolor}
\usepackage{newtxtext,newtxmath}
\usepackage[normalem]{ulem}
\usepackage{diagbox}
\usepackage{array, makecell}
\usepackage{xspace}
\usepackage{hyperref}
\usepackage{soul}
\hypersetup{colorlinks=true, citecolor=blue, urlcolor=blue, linkcolor=blue}
\newcommand\nd{\textsuperscript{nd}\xspace}

\newcommand{\edit}[1]{#1}

\renewcommand{\arraystretch}{1.4}
\begin{document}
\newcommand{\euclidemutwo}{\textsc{EuclidEmulator2}}
\newcommand{\eetwo}{\textsc{EE2}}

\title{Modeling nonlinear scales for dynamical dark energy cosmologies with COLA}

\author{João Rebouças}
\affiliation{CBPF - Brazilian Center for Research in Physics, Xavier Sigaud st. 150, zip 22290-180, Rio de Janeiro, RJ, Brazil}
\affiliation{Instituto de Física Teórica da Universidade Estadual Paulista, R. Dr. Bento Teobaldo Ferraz, 271, Bloco II, Barra-Funda - São Paulo/SP, Brazil}

\author{Victoria Lloyd}
\affiliation{Department of Physics \& Astronomy, Stony Brook University, Stony Brook, NY 11794, USA}

\author{Jonathan Gordon}
\affiliation{Department of Physics \& Astronomy, Stony Brook University, Stony Brook, NY 11794, USA}
\affiliation{C. N. Yang Institute for Theoretical Physics, Stony Brook University, Stony Brook, NY, 11794, USA}

\author{Guilherme Brando}
\affiliation{CBPF - Brazilian Center for Research in Physics, Xavier Sigaud st. 150, zip 22290-180, Rio de Janeiro, RJ, Brazil}
\author{Vivian Miranda}
\affiliation{C. N. Yang Institute for Theoretical Physics, Stony Brook University, Stony Brook, NY, 11794, USA}
\date{\today}

\begin{abstract}
Upcoming galaxy surveys will bring a wealth of information about the clustering of matter at small scales, but modeling small-scale structure beyond $\Lambda\mathrm{CDM}$ remains computationally challenging. While accurate $N$-body emulators exist to model the matter power spectrum for $\Lambda\mathrm{CDM}$ and some limited extensions, it's unfeasible to generate $N$-body simulation suites for all candidate models. Motivated by recent hints of an evolving dark energy equation of state from galaxy surveys, we assess the viability of employing the COmoving Lagrangian Acceleration (COLA) method to generate simulation suites assuming the $w_0w_a$ dynamical dark energy model. Following up on our previous work, we combine COLA simulations with an existing high-precision $\Lambda\mathrm{CDM}$ emulator to extend its predictions into new regions of parameter space. We assess the precision of our emulator at the level of the matter power spectrum, finding that our emulator can reproduce the nonlinear boosts from \euclidemutwo~ at less than $2\%$ error. Moreover, we perform an analysis of a simulated cosmic shear survey akin to future data from the Legacy Survey of Space and Time (LSST) first year of observations, assessing the differences in parameter constraints between our COLA-based emulator and the benchmark emulator. We find our emulator to be in excellent agreement with the benchmark, achieving less than $0.3\sigma$ shifts in cosmological parameter constraints across multiple fiducial cosmologies. We further compare our emulator's performance to a commonly used approach: assuming the $\Lambda\mathrm{CDM}$ boost can be employed for extended parameter spaces without modification. We find that our emulator yields a significantly smaller $\Delta\chi^2$ distribution, parameter constraint biases, and a more accurate figure of merit compared to this second approach. These results demonstrate that COLA emulators provide a computationally efficient and physically motivated path forward for modeling nonlinear structure in extended cosmologies, offering a practical alternative to full $N$-body suites in the era of precision cosmology.
\end{abstract}

\maketitle

\section{Introduction}
\label{sec:Introduction}
In recent decades, galaxy surveys have been able to map the large-scale structure of the Universe with great precision, being as competitive as CMB surveys. Current photometric surveys, such as DES~\cite{desy1_shear, desy1_3x2, des_y1_extensions, desy3_shear1, desy3_shear2, desy3_cov, desy3_maglim, desy3_redmagic, desy3_3x2, des_y3_extensions}, KiDS~\cite{kids-legacy, kids-legacy-calibration, kids1000_bgs, kids_euclid_pseudocl, kids1000_density_split, kids1000_enhance_cal, kids1000_extensions, kids1000_halo_constraints, kids-1000, kids1000_ia_lrg, kids1000_photoz_cal, kids1000_qso, kids1000_x_cmb}, HSC~\cite{hsc_y1, hsc_y1_2pt, hsc_overview, hsc_photoz_cal, hsc_shear_cal, hsc_y3}, and spectroscopic surveys, such as DESI~\cite{desi_dr1, desi_dr2_validation, desi_dr2_validation_lya, desi_dr2_results_lya, desi_dr2_bao, desi_dr2_extensions, desi_dr2_neutrinos}, BOSS/eBOSS~\cite{boss_bao, boss_bao_harmonic, boss_results, boss_rsd, boss_rsd_harmonic, eboss_overview, eboss_lrg, eboss_forecast, eboss_photoz_des, eboss_qso_bao, eboss_qso_clustering, eboss_qso_selection, eboss_rsd_qso, eboss_rsd_qso_2}, and WiggleZ~\cite{wigglez_bao, wigglez_bao_2, wigglez_blue_gals, wigglez_results, wigglez_rsd}, have placed percent-level constraints on $\Lambda$CDM parameters, demonstrating the success of the theoretical model in describing independent datasets. However, in recent years, tensions in the cosmological parameter constraints have begun to arise with the increase in precision. Notably, most galaxy surveys report a mildly lower value for the structure growth parameter $S_8$ than those from CMB measurements (see \textit{e.g.} \cite{cosmology_intertwined_s8}). Moreover, recent findings from the DESI collaboration, as well as Type Ia supernovae datasets such as Pantheon+, Union3, and DESY5~\cite{union3, desy5, desi_dr1, desi_dr2_bao, desi_dr2_extensions, joao_desi_de} favor an evolution of the dark energy equation of state. Whether these results truly indicate new physics is still under debate \cite{ROY2025101912,PhysRevD.111.023512,chakraborty2025desiresultshintcoupled,huang2025desidr1dr2evidencedynamical,chaudhary2025doesdesidr2challenge}.


Forthcoming Stage-IV galaxy surveys, such as the Vera Rubin Observatory's LSST \cite{lsst-survey-specifications}, Euclid \cite{euclid_overview}, SphereX \cite{spherex}, and Roman \cite{roman_multiprobe}, will enable higher-precision measurements, especially at small scales where non-linearities in the matter density field become increasingly sizable \cite{lacasa2022}, and will be decisive for investigating dark energy dynamics. Analyses of galaxy surveys rely on a key theoretical prediction: the matter power spectrum, as galaxies are biased tracers of the underlying matter density field. At large scales and early times, the power spectrum can be quickly and accurately computed using Einstein-Boltzmann solvers such as \textsc{camb}~\cite{camb} and \textsc{class}~\cite{class, class2}. However, for small scales and low redshifts, linear perturbation theory breaks down, and accurately modeling non-linearities becomes a task of central importance.


Although \textit{N}-body simulations offer the most accurate predictions in the nonlinear regime, they are computationally expensive --- each one demanding tens of thousands of CPU hours~\cite{Schneider:2015yka, pkdgrav3, gadget4}. As a result, integrating simulations into Bayesian analyses is a prohibitive task, as theoretical predictions must be provided for $\mathcal{O}(10^5-10^6)$ points in the parameter space. To address this issue, machine learning emulators trained on \textit{N}-body simulations have been developed for $\Lambda$CDM and simple, widely adopted extensions such as the phenomenological $w_0w_a$CDM~\cite{cpl, cpl_2} parametrization for dark energy, as well as massive neutrinos with their total mass as a free parameter. Examples include \euclidemutwo~\cite{euclidemu2}, \textsc{bacco}~\cite{bacco}, \textsc{CosmicEmu} \cite{Lawrence_2017}, the Dark Quest emulator \cite{DarkQuest2025, DarkQuest2022}, the CSST Emulator~\cite{Chen_2025, chen2025csstii, zhou2025csstiii}, \textsc{aemulus}~\cite{DeRose:2023dmk, DeRose:2018xdj, McClintock:2018uyf, Zhai:2018plk}, among others. At the same time, due to the sheer amount of candidate cosmological models and the high cost of running \textit{N}-body simulations, emulators for broader extensions of $\Lambda$CDM are still scarce: examples are emulators to specific modified gravity theories (\textit{e.g.}~\cite{cola_mg, axion_nonlin, emantis, forge}) where we have full \textit{N}-Body simulations~\cite{mg_nbody_comparison}, as well as some hydrodynamical emulators.

A viable alternative to using full $N$-body simulations for constructing matter power spectrum emulators is to use well-established approximate methods for extended cosmological models~\cite{McEwen:2016fjn, Chudaykin:2020aoj, DAmico:2020kxu, cobra, sym-eft}. These methods reduce the computational complexity of running hundreds of simulations to train emulators, at the cost of losing accuracy in deep non-linear scales. One compelling approach is to use the COmoving Lagrangian Acceleration (COLA) method, which combines Lagrangian Perturbation Theory with a particle-mesh (PM)~\cite{PM1988} evolution scheme to approximate \textit{N}-body results while being cheaper than the usual \textit{N}-body methods by 1-2 orders of magnitude \cite{Tassev:2013pn, cola_mock_galaxy_catalogs}. Emulators created using pure COLA simulations are prone to small-scale inaccuracies when compared directly to their \textit{N}-body counterparts. To mitigate this effect, the work of~\cite{Brando:2022gvg} introduces an approach that combines COLA simulations with predictions from high-accuracy $\Lambda$CDM emulators or full \textit{N}-body results, leveraging COLA's reduced computational cost and dramatically increasing small-scale accuracy simultaneously. Our previous work~\cite{cola_wcdm} has validated this approach, creating an emulator for the matter power spectrum of COLA simulations under the $w$CDM cosmological model, and testing it in a mock Stage-IV cosmic shear analysis. As such, we aim to demonstrate here that the hybrid approach of COLA-based emulators combined with high-resolution $\Lambda$CDM emulators can provide unbiased cosmological parameter constraints when compared to full \textit{N}-body methods.


In this work, we now present a final validation of our COLA-based emulators on the $w_{0}w_{a}$CDM cosmological model, where the dark energy equation of state evolves linearly with the scale factor. This parametrization represents the most widely used and general extension of $\Lambda$CDM for which high-accuracy emulators currently exist, and remains central to ongoing investigations into dynamical dark energy~\cite{desi_dr1, desi_dr2_bao}. We extend our machine learning pipeline, combining COLA simulations with $\Lambda$CDM emulators, to predict the nonlinear matter spectrum across the $w_0w_a$CDM parameter space. We train a simple neural network to emulate the nonlinear correction factor (\textit{i.e} the boost) from COLA, correcting for small-scale inaccuracies by referencing boosts from a high-fidelity $\Lambda$CDM emulator. This hybrid approach enables fast predictions across an extended cosmological parameter space while maintaining consistency with $N$-body precision.
To validate our emulator in a cosmological inference setting, we perform a simulated cosmic shear analysis using survey specifications consistent with LSST's first year of observations (LSST-Y1) \cite{lsst-survey-specifications}. We compare parameter constraints derived using both our pipeline and a benchmark $N$-body emulator, chosen as \euclidemutwo, quantifying their disagreement with standard tension metrics \cite{Mortonson_2010, Trotta_2008, Shapiro_2009}.

Additionally, we also benchmark our emulator against a widely used approximation method in beyond-$\Lambda\mathrm{CDM}$ analyses for models without dedicated nonlinear simulations~\cite{detg_model, ede_act, ide_1, ide_2, ide_3, ide_4}: projecting the nonlinear boost from the nearest $\Lambda\mathrm{CDM}$ cosmology. This projection method assumes that nonlinear corrections calibrated in $\Lambda\mathrm{CDM}$ remain valid in nearby extended cosmologies, providing a computationally inexpensive workaround but at the cost of uncontrolled systematics. In contrast, we demonstrate that for dynamical dark energy models, we find that the $\Lambda$CDM projection approach may introduce significant deviations in both goodness-of-fit and parameter constraints. Meanwhile, our COLA-based emulator reproduces the predictions of high-precision $N$-body emulators without bias.

This paper is organized as follows: Section~\ref{sec:Methodology} describes the COLA simulations, cosmological parameters, simulation output processing, emulator construction, and validation; in Section~\ref{sec:lsst}, we present our LSST-Y1 simulated cosmic shear analysis and the tension metrics used to assess their differences; in Section~\ref{sec:Results}, we present and discuss the results of the LSST-Y1 simulated analysis; finally, we conclude in Section~\ref{sec:Conclusion}.

\section{COLA Emulator}
\label{sec:Methodology}

\subsection{COLA Simulation Suite}
\label{sec:COLA}

The COmoving Lagrangian Acceleration (COLA) algorithm \cite{Tassev:2013pn} is a fast approximate method for $N$-body simulations, wherein particles evolve in a frame comoving with trajectories calculated using Lagrangian Perturbation Theory (LPT), most commonly 2\nd order Lagrangian perturbation theory (2LPT). For small scales, the method computes the force by using a Particle-Mesh (PM) algorithm, where the residual displacements not captured by LPT are added to the trajectories. 

COLA has been shown to agree with full $N$-body simulations, at the level of the power spectrum at up to $k\sim 1h/$Mpc, as well as when predicting ratios of the modified gravity power spectrum and the $\Lambda$CDM one, the so-called boost function in modified gravity~\cite{Fiorini:2023fjl,Brando:2022gvg,Izard:2015dja,hicola}. Despite being 1-2 orders of magnitude faster than a full $N$-body run, the computational cost of these approximations is still too high for direct use in the $\mathcal{O}(10^6)$ computations of the matter power spectrum required for Monte Carlo searches. A practical alternative is to use a fixed set of COLA simulations to train emulators for the matter power spectrum, enabling efficient interpolation across cosmological parameter space. Our previous work demonstrated this approach for $w$CDM \cite{cola_wcdm}; here, we extend it to $w_0w_a$CDM and evaluate its performance relative to the benchmark \euclidemutwo~\cite{euclidemu2}, which achieves $\lesssim 1\%$ precision for $w_0w_a\mathrm{CDM} + \sum m_\nu$ up to $k = 10 \; h/\mathrm{Mpc}^{-1}$ and $z \leq 3$. 

\subsubsection{Simulation Settings}\label{sec:SimSettings}

We use the COLA algorithm as implemented in the public \textsc{fml}\footnote{\url{https://github.com/HAWinther/FML}} code. Each simulation is performed in a box of size $L = 1024 h^{-1}\mathrm{Mpc}$, populated with $N_{\mathrm{part}} = 1024^3$ particles, initialized at $z_\mathrm{ini} = 19$, and evolved over 51 time steps chosen to maintain a uniform time resolution of $\Delta a \approx 0.02$. The force grid uses $N_{\mathrm{mesh}} = 2048^3$ cells, and the power spectra are calculated on-the-fly using a $N_\mathrm{pk-mesh}^3 = 1024^3$ grid. Therefore, the corresponding Nyquist frequency is $k_\mathrm{Nyq} = \pi N_\mathrm{pk-mesh}/L = \pi \; h/\mathrm{Mpc}$. To avoid aliasing, we restrict our analysis to $k \leq k_\mathrm{Nyq}$~\cite{estimating_pk_sims}. Our choices are based on Reference~\cite{Brando:2022gvg} and are validated therein.

Initial conditions are generated using 2LPT, and we employ the forward approach~\cite{Angulo_2022} for our simulations. We provide the linear transfer functions of matter density, velocity, and relativistic species' densities at each time step using \textsc{class}\footnote{\url{https://github.com/lesgourg/class_public}} in synchronous gauge, and convert to the $N$-body gauge~\cite{Fidler:2015npa,Fidler:2017ebh,Tram:2018znz,Brando:2020ouk} in COLA. Our simulations were run in the Seawulf\footnote{\url{https://rci.stonybrook.edu/HPC}} cluster. With these settings, and using 128 cores, one COLA $w_0w_a$ simulation takes approximately 40 minutes to finish, and requires a total RAM of approximately 950 GB.

To suppress sample variance from finite box effects at large scales ($k \approx 1/L$), we use the pairing-and-fixing method~\cite{Angulo_2016}, in which we generate Gaussian random field modes with a fixed amplitude, $\delta_{i,\mathrm{lin}}$, but with phase shifts of $\pi$ with respect to one another. The initial overdensity fields are sampled as
\begin{equation}
    \delta_{i,\mathrm{lin}} = \sqrt{P_i}e^{\mathrm{i}\theta_i}.
\end{equation}
where $\theta^i$ is a random phase and $P_i$ the initial power spectrum. Averaging over each pair, we find that the result substantially suppresses the effects of cosmic variance. This strategy was chosen for this work following~\cite{cola_wcdm} and \cite{euclidemu2}.   

\subsubsection{Definition of the Parameter Space}\label{sec:ParamSpace}

We consider the cosmological $w_0w_a$CDM model, where the dark energy equation of state is parametrized as
\begin{align}
    w(a) = w_0 + w_a(1-a), \label{eq:deEoS}
\end{align}
with $a$ being the scale factor, and $w_0$ and $w_a$ control the present-day value and time derivative of the dark energy equation of state, respectively. The $\Lambda$CDM model is recovered in the limit $w_0 = -1$ and $w_a = 0$. The free cosmological parameters are:
\begin{itemize}
    \item $\Omega_m$, the total baryon density, 
    \item $\Omega_b$, the total matter density,
    \item $h = H_0/(100 \, \mathrm{km \, s^{-1} \, Mpc^{-1}})$, the dimensionless Hubble parameter,
    \item $A_s$, the amplitude of initial scalar fluctuations, 
    \item $n_s$, the scalar spectral index,
    \item $w_0$ and $w_a$, the dark energy equation of state parameters.
\end{itemize}
We fix the summed neutrino masses to the minimum value allowed by neutrino oscillation experiments, $\sum m_\nu = 0.058$ eV, assuming three degenerate massive species \cite{PhysRevD.95.096014,PhysRevD.98.030001}. \edit{While other emulators available in the literature are valid for varying $\sum m_\nu$, for now we restrict ourselves to this simpler scenario. We remark that COLA has already been validated to model massive neutrinos with good accuracy~\cite{Euclid:2022qde}, and our methodology can naturally include the neutrino mass sum $\sum m_\nu$ as an additional parameter. }

The parameter space boundaries are described in Table \ref{tab:param_space}, set to match those of \euclidemutwo, which we have adopted as our benchmark for comparison. We emphasize that this choice is arbitrary, and our methodology is agnostic to the benchmark we have chosen. To improve model performance near parameter space edges, our COLA training simulations are sampled from an expanded box where each parameter interval has been stretched symmetrically by 10\%. Cosmologies within this volume are selected using Latin hypercube sampling to ensure uniform coverage for training and validation.

\begin{table}[t]
\centering
\renewcommand{\arraystretch}{1.5}
\begin{tabular}{|@{\hskip 5pt} c @{\hskip 5pt}|@{\hskip 5pt} c @{\hskip 5pt}|@{\hskip 5pt} c @{\hskip 5pt}|@{\hskip 5pt} c @{\hskip 5pt}|@{\hskip 3pt} c @{\hskip 3pt}|@{\hskip 5pt} c @{\hskip 5pt}|@{\hskip 5pt} c @{\hskip 5pt}|@{\hskip 5pt} c @{\hskip 5pt}|}
\hline
 & $\Omega_m$ & $\Omega_b$ & $n_s$ & $A_s\times10^{9}$ & $h$ & $w_0$ & $w_a$ \\ \hline
Min & $0.24$ & $0.04$ & $0.92$ & $1.7 $ & $0.61$ & $-1.3$ & $-0.7$ \\ \hline
Max & $0.40$ & $0.06$ & $1.00$ & $2.5 $ & $0.73$ & $-0.7$ & $0.5$ \\ \hline
Center & $0.319$ & $0.05$ & $0.96$ & $2.1 $ & $0.67$ & $-1$ & $0$ \\ \hline
\end{tabular}
    \caption{Parameter space validity bounds of our COLA-based emulator. The training set is drawn from a slightly bigger hypercube, where each dimension is stretched by 10\% in each direction (\textit{e.g.} $ 0.224 < \Omega_m < 0.416 $). We also define a center cosmology chosen to agree with the \euclidemutwo~reference cosmology~\cite{euclidemu2}.}
\label{tab:param_space}
\end{table}

\subsection{Emulation of COLA Boosts}
\subsubsection{Emulator Prototypes with \textsc{halofit}}
\label{sec:halofit}

Our goal is for the emulation error (\textit{i.e.}, the error in recovering COLA boosts from a predetermined test set excluded from training) to be significantly smaller than the intrinsic COLA approximation error relative to full $N$-body simulations. To determine optimal hyperparameters, such as training set size and emulator architecture, we perform mock tests using \textsc{halofit}~\cite{Takahashi:2012em} boosts. We generated training datasets with $N_\mathrm{train} \in [500, 600, 700, 800, 1000]$ \textsc{halofit} boosts and a test dataset with $N_\mathrm{test} = 200$. We found that 600 simulations were sufficient to achieve $\sim 0.1\%$ error at $k = 1 \; h/\mathrm{Mpc}$; conservatively, we adopt $N_\mathrm{train} = 700$ and $N_\mathrm{test} = 200$ for our COLA emulator.

\subsubsection{Post-processing the Simulation Boosts}
We define the nonlinear boost as
\begin{equation}\label{eq:boost}
    B^\mathrm{X}(k,z|\boldsymbol{\theta}) \equiv \frac{P^\mathrm{X}(k,z|\boldsymbol{\theta})}{P^{\mathrm{L}}(k,z|\boldsymbol{\theta})},
\end{equation}
where $\boldsymbol{\theta}$ refers to a point in the $w_0w_a$CDM parameter space, $P^X(k,z|\boldsymbol{\theta})$ is the matter power spectrum for cosmology $\boldsymbol{\theta}$, either linear (denoted $P^\mathrm{L}$), or calculated using COLA or another \textit{N}-body method (generically denoted $P^\mathrm{X}$). Prior to computing $B^\mathrm{COLA}$, we subtract the shot noise power spectrum, $P_\mathrm{SN} = (L/N_\mathrm{part})^3 = 1 (\mathrm{Mpc}/h)^3$, from $P^\mathrm{COLA}$. At high redshift, $z > 1.182$, aliasing of the $k$ modes near the Nyquist frequency leads to a power spectrum less than the shot noise for some simulations, and the subtraction would lead to unphysical negative values \cite{Angulo_2022}; for these redshifts, we choose to cut the scales at half of the Nyquist frequency, $k^{z > 1.182} \leq (\pi/2) \; h/\mathrm{Mpc}$, following our procedure in~\cite{cola_wcdm} (also see~\cite{estimating_pk_sims}).

We then perform several transformations to optimize the inputs and outputs of our emulator. For instance, machine learning techniques are known to perform poorly if the features span several orders of magnitude. To stabilize the following procedures, we normalize the cosmological parameters $\boldsymbol{\theta}$ to $[-1, 1]$ according to
\begin{equation}\label{eq:params_norm}
    \boldsymbol{\theta}_N = -1 + 2 \frac{\boldsymbol{\theta} - \boldsymbol{\theta}_\mathrm{min}}{\boldsymbol{\theta}_\mathrm{max} - \boldsymbol{\theta}_\mathrm{min}},
\end{equation}
where minimum and maximum values correspond to the training set boundaries, \textit{i.e.}, stretching the intervals of Table~\ref{tab:param_space} by $10\%$ in each direction. Furthermore, we standardize the boosts using
\begin{equation}\label{eq:boost_norm}
    B_N^\mathrm{COLA}(k,z|\boldsymbol{\theta}) = \frac{B^\mathrm{COLA}(k,z|\boldsymbol{\theta}) - \bar{B}^\mathrm{COLA}(k,z)}{\sigma_B(k, z)},
\end{equation}
where $B_N^\mathrm{COLA}(k, z | \boldsymbol\theta)$ is the normalized COLA boost, $\bar{B}(k, z)$ is the average of all boosts in the training set, and $\sigma_B(k, z)$ is their standard deviation. We then perform a Principal Component Analysis (PCA) decomposition of the COLA boosts using \textsc{scikit-learn}~\cite{scikit-learn} to reduce dimensionality. We retain  $N_\mathrm{PC} = 15$ components, which are sufficient to recover the test set boosts to within $0.2\%$. 


\subsubsection{Neural Network Emulator}
After post-processing, we train our emulator with the normalized cosmological parameters as input features and the principal components as targets. We use a fully connected neural network with three hidden layers, each with  1024 neurons, with a mean squared error loss function,
\begin{equation}
    \mathcal{L} = \sum_{i = 1}^{N_\mathrm{train}}\sum_{j = 1}^{N_\mathrm{PC}}(\alpha_j^{i, \mathrm{train}} - \alpha_j^{i, \mathrm{pred}})^2,
\end{equation}
where $\alpha_j^{i, \mathrm{train}}$ is the $j$-th principal component coefficient of the $i$-th cosmology in the training set, and $\alpha_j^{i, \mathrm{pred}}$ the corresponding prediction. We use the parametric activation function~\cite{cosmopower, speculator}
\begin{equation}
    y^{m+1}_n = \left[\gamma^m_n + (1 - \gamma^m_n)\frac{1}{1 + e^{-\beta^m_n y^m_n}}\right]\tilde{y}^m_n,
    \label{eq:nn_activation}
\end{equation}
where $y^{m+1}_n$ is the value of the $n$-th neuron of the $(m+1)$-th layer, $\tilde{y}^m_n$  the $n$-th neuron from the $(m+1)$-th layer after the application of weights and biases, and $\gamma^m_n$ and $\beta^m_n$ are parameters of the activation function that can be back-propagated during training. We use the \textsc{Adam}~\cite{adam_opt} optimizer to train the model parameters.

\subsubsection{Boost Errors}
\label{sec:evaluating_emulator}
\begin{figure}
    \centering
    \includegraphics[width=0.95\linewidth]{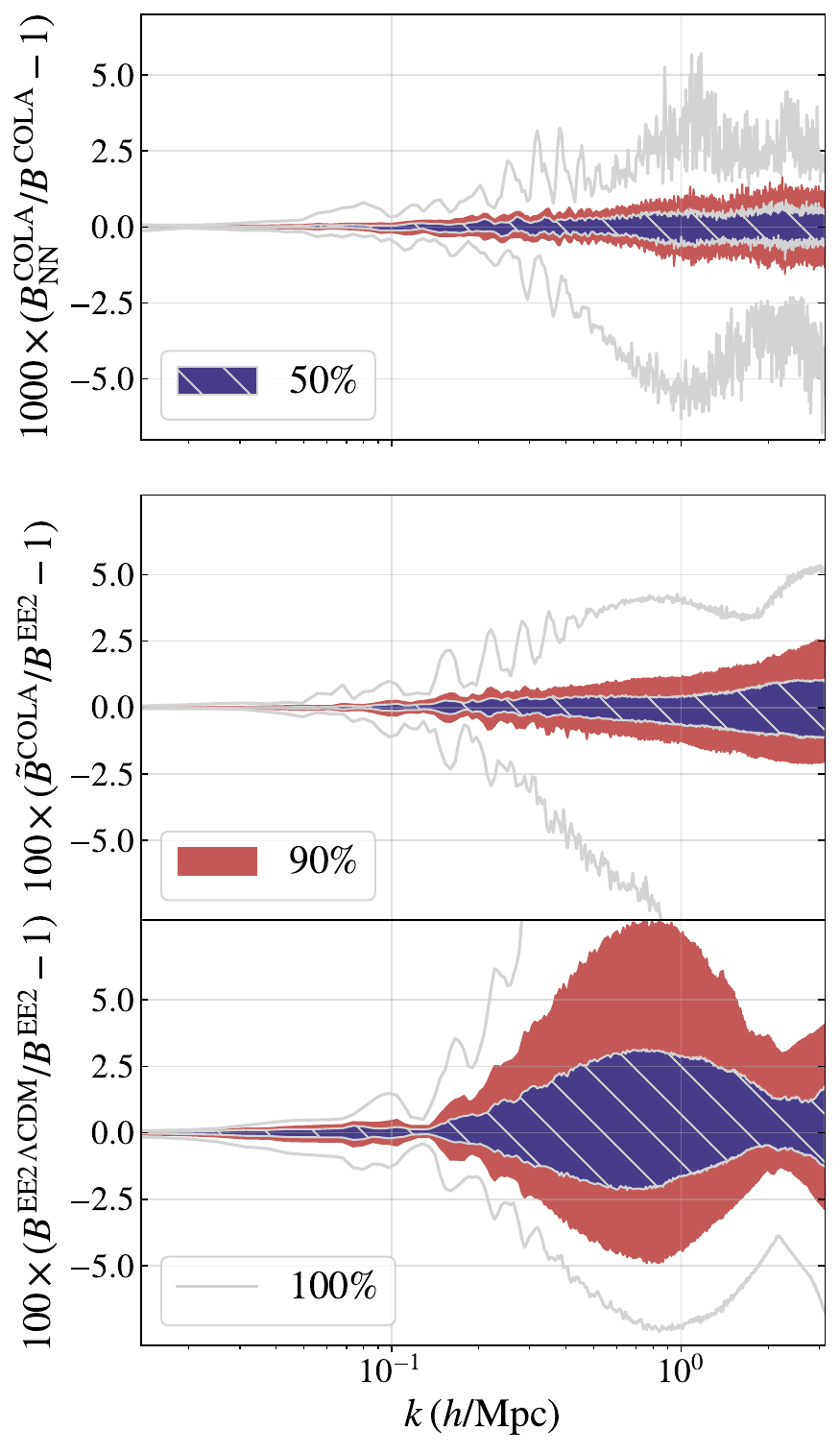}
    \caption{\textbf{From top to bottom:} \textbf{1)} Relative errors between the COLA boosts $B^\mathrm{COLA}$ predicted by the emulator versus those obtained from the test set simulations. \textbf{2)} Relative errors between $\tilde{B}^\mathrm{COLA}$ (see Equation~\ref{eq:inf_refs}) and the boosts from \eetwo. \textbf{3)} Relative errors between $B^{\mathrm{EE2} \; \Lambda \mathrm{CDM}}$ and \eetwo. Colors in all panels denote the percentile of cosmologies around the mean: blue contours enclose $50\%$ of cosmologies, red contours enclose $90\%$ of cosmologies, and the outer gray lines enclose $100\%$ of cosmologies. All panels show results for $z = 0$, see Appendix~\ref{app:errors_higher_z} for the equivalent plots at higher redshifts.}
    \label{fig:errors}
\end{figure}

We perform a series of accuracy checks on the emulator outputs. First, to assess the accuracy of our neural network, we compare the emulator's predictions for test set cosmologies, unseen in the training procedure, against the actual COLA simulations. The relative errors are shown in the first panel of Figure~\ref{fig:errors}. At $k = 1\, h/\mathrm{Mpc}$, $90\%$ of the test set cosmologies have an emulation error within $0.1\%$. Comparing these direct emulation errors to the errors on the corrected boosts $\tilde{B}^\mathrm{COLA}(k,z)$ (second panel), we note that the errors between the two differ by an order of magnitude. This indicates that, in the context of comparing COLA with high-precision simulations, the COLA emulator faithfully reproduces its simulations, and differences between the emulators can be attributed to the COLA approximation rather than the performance of the machine learning model.

As per \cite{Brando:2022gvg}, COLA simulations increasingly lose power\footnote{This loss of power is well known to PM \textit{N}-body codes, which fail to resolve the internal dynamics of halos. This trend of losing power starts roughly at a scale at which the pure $1$-halo term of the halo model would dominate the power spectrum.} at progressively smaller scales, leading to typical errors of 10\% at $k=1\,  h/\mathrm{Mpc}$ for raw COLA boosts $B^{\mathrm{COLA}}$, as defined in Equation \ref{eq:boost}. However, this power loss is cosmology-independent. Our previous work~\cite{cola_wcdm} showed that the best technique to build a COLA-based emulator is to leverage existing high-precision emulators in $\Lambda$CDM, using COLA only to extend the results into new dimensions, \textit{i.e.}, extra model parameters. This idea is encoded in the following expression for the nonlinear boost,
\begin{equation}\label{eq:inf_refs}
    \tilde{B}^\mathrm{COLA}(k, z | \boldsymbol{\theta}) = B^\text{N-body}(k, z | \boldsymbol{\theta}_p) \times \frac{B^\mathrm{COLA}(k, z | \boldsymbol{\theta})}{B^\mathrm{COLA}(k, z | \boldsymbol{\theta}_p)},
\end{equation}
where $\boldsymbol{\theta}_p$ is the projection of $\boldsymbol{\theta}$ in the $\Lambda$CDM subspace and $B^\mathrm{N-body}$ is the nonlinear boost obtained from our benchmark $N$-body prescription. We choose \euclidemutwo~as the base N-body prescription.

The second panel of Figure~\ref{fig:errors} shows the relative difference between $\tilde{B}(k, z)$, calculated using Equation \ref{eq:inf_refs}, and the benchmark emulator predictions, $B^\mathrm{EE2}(k, z)$, for all test set cosmologies. At $k = 1\, h/\mathrm{Mpc}$, $90\%$ of cosmologies have emulation errors within $2\%$, with $50\%$ of cosmologies contained well within $1\%$. This demonstrates that our method successfully mitigates the accumulation of errors typical of COLA simulations in the nonlinear regime, allowing us to generate accurate predictions across our target $k$ range. Furthermore, the cosmologies with larger errors are those with higher values of $w_0+w_a$, a region of the parameter space excluded by current data. 


Finally, we consider a third nonlinear prescription: using the nonlinear boost from \euclidemutwo~in the $\Lambda\mathrm{CDM}$ subspace; this approach will be denoted as \eetwo~$\Lambda$CDM. For this purpose, we compute the nonlinear boosts for $w_0w_a$CDM cosmologies using \euclidemutwo, setting $w_0 = -1$ and $w_a = 0$. The third panel of Figure~\ref{fig:errors} shows relative errors between $B^{\mathrm{EE2} \, \Lambda \mathrm{CDM}}$ and the actual \euclidemutwo~boosts. The $90\%$ percentile shows errors of the order of $7.5\%$ at $k = 1 \; h/\mathrm{Mpc}$, significantly worse than the COLA errors of the panel above.

These results support the viability of our emulator for parameter inference in $w_0w_a\mathrm{CDM}$, as the emulated boost $\tilde{B}(k,z)$ agrees with our $N$-body proxy at a level suitable for upcoming precision cosmology experiments \cite{Schneider_2016} while requiring significantly less computational expense compared to traditional $N$-body methods. In the following, we investigate how the differences shown in Figure~\ref{fig:errors} impact the parameter constraints from simulated cosmic shear analysis.

\section{Analysis of LSST-Y1 Simulated Data}
\label{sec:lsst}

\subsection{Simulating Cosmic Shear Data}

We simulate cosmic shear observations based on LSST-Y1, following the methodology of~\cite{cosmocov, accelerated-inference-emulator-lsst} and detailed in~\cite{cola_wcdm}. Survey specifications, source galaxy redshift distributions, and nuisance parameter priors are taken from the LSST DESC Science Requirements Document~\cite{lsst-survey-specifications}, and summarized in Table~\ref{tab:lsst_specifications}. The redshift distribution is modeled as a Smail distribution convolved with a Gaussian uncertainty $0.02(1+z)$ and divided into five tomographic bins with equal galaxy number densities.

\begin{table}[t]
	\centering
	\begin{tabular}{| lcc|} 
        \hline
        \textbf{Parameter} & \textbf{Fiducial} & \textbf{Prior}\\ \hline
        \textbf{Survey specifications} & & \\
        Area & $12300 \; \mathrm{deg}^2$ & --\\
        Shape noise per component & $0.26$ & --\\
        $n_\mathrm{eff}^\mathrm{sources}$ & $11.2 \; \mathrm{arcmin}^{-2}$ & -- \\
        \hline
        \textbf{Photometric redshift offsets} & & \\
        $\Delta z_{\mathrm{source}}^{i}$ & 0 & $\mathcal{N}$[0, 0.002]\\
        \hline
        \textbf{Intrinsic alignment (NLA)}& &\\
        $a_{1}$ & 0.7 & $\mathcal{U}$[-5, 5]\\
        $\eta_{1}$ & -1.7 &  $\mathcal{U}$[-5, 5]\\
        \hline
        \textbf{Shear calibration} & & \\
        $m^i$ & 0 & $\mathcal{N}$[0, 0.005]\\
        \hline
	\end{tabular}
    \caption{Mock survey specifications for our simulated analysis, and nuisance parameter priors. $\mathcal{U}[a,b]$ denotes an uniform distribution with edges $[a,b]$, while $\mathcal{N}[a,b]$ denotes a Gaussian distribution with mean $a$ and standard deviation $b$. Tomographic bin indices are denoted by $i$, and all our priors are the same for all bins.}
    \label{tab:lsst_specifications}
\end{table}
The cosmic shear two-point correlation functions $\xi^{ij}_\pm(\theta)$ are computed by first evaluating in Fourier space, $C^{ij}_{\kappa\kappa}(\ell)$, using the nonlinear matter power spectrum via the Limber approximation, then transforming to real space via the analytic functions in Appendix A of~\cite{desy3_cov}. We compute $\xi^{ij}_\pm$ in 26 logarithmically spaced angular bins between 2.5 and 900 arcmin, averaging over each bin. We include standard self-calibrating systematics in our computation of $\xi_\pm$ — photometric redshift uncertainties, multiplicative shear calibration, and the non-linear alignment (NLA) model of intrinsic galaxy alignments (see, \textit{e.g.}, \cite{ia-overview, ia-review}).

Likelihood analyses are performed using \textsc{Cocoa}, the \textsc{Cobaya}-\textsc{CosmoLike} Joint Architecture\footnote{\url{https://github.com/CosmoLike/cocoa}}~\cite{Krause:2016jvl, Lewis:2013hha, Torrado:2020dgo}. Linear power spectra are computed with \textsc{camb}~\cite{Lewis:1999bs,Howlett:2012mh}, and nonlinear corrections are applied using either $\tilde{B}$ (Eq.~\ref{eq:inf_refs}), $B^{\eetwo~\Lambda\mathrm{CDM}}$, or \euclidemutwo. We use MCMC sampling to explore the parameter space and assess convergence using the Gelman–Rubin criterion ($|R - 1| < 0.01$) \cite{gelman_rubin}.

To evaluate the emulator's accuracy in the parameter space, we define 29 fiducial cosmologies within the emulator's validity range. One data vector is generated at the center cosmology shown in Table~\ref{tab:param_space}. We then vary cosmological parameters to their intermediate "low" ($\downarrow$) and "high" ($\uparrow$) values, shown in Table~\ref{tab:fiducials}. We define four cosmologies by varying both $w_0$ and $w_a$ for each combination of their low and high values, keeping other parameters fixed at their central values. We further define 24 fiducial cosmologies varying $w_0$, $w_a$, and one of $\Omega_m$, $A_s$, or $n_s$, using all combinations of their low and high values. All data vectors are generated using \euclidemutwo~as the nonlinear prescription, and the covariance is computed analytically with \textsc{CosmoCov} \cite{cosmocov}. We do not include a Gaussian noise realization in the fiducial data vectors.


\begin{table}[t]
\centering
\renewcommand{\arraystretch}{1.4}
\begin{tabular}{| c || >{\hspace{4pt}}c<{\hspace{4pt}} | >{\hspace{4pt}}c<{\hspace{4pt}} | >{\hspace{4pt}}c<{\hspace{4pt}} | >{\hspace{4pt}}c<{\hspace{4pt}} | >{\hspace{4pt}}c<{\hspace{4pt}} |}
\hline
$\theta$ & $\Omega_m$ & $10^{9}A_s$  & $n_s$ & $w_0$ & $w_a$ \\
\hline\hline
$\theta^{\uparrow}$ & 0.36 & 2.3 & 0.98 & -0.85 & 0.25 \\
$\theta^{\downarrow}$ & 0.28 & 1.9 & 0.94 & -1.15 & -0.35 \\
\hline
\end{tabular}
\caption{"High" ($\uparrow$) and "low" ($\downarrow$) cosmological parameter values used to construct the fiducial cosmologies of our analyses. In this notation, fiducial cosmologies are labeled in the text by the parameters shifted from the central values listed in Table~\ref{tab:param_space}.}
\label{tab:fiducials}
\end{table}
To mitigate biases from emulator inaccuracies on small scales, which may degrade the goodness-of-fit, we apply three angular scale cuts (C1-C3) following \cite{cola_wcdm}, removing $\xi_\pm$ measurements below a minimum angular separation, $\theta_\mathrm{min}$. The corresponding wavenumbers are shown in Table \ref{tab:k_cuts}; cuts are more aggressive for $\xi_-$ due to its sensitivity to smaller scales.


Finally, to compute the cosmic shear integrals beyond the emulator’s $k$-range, we extrapolate $\log(\tilde{B}(k))$ versus $\log(k)$ using a linear fit for both COLA and \eetwo~emulators. A Savitsky–Golay filter (order 1, window length 5) is applied to the last entries of the $\tilde{B}^\mathrm{COLA}$ vector to suppress noise before extrapolation.



\begin{table}[t]
\centering
\begin{tabular}{| c | c | c | c | c | c |}
\hline
\rule{0pt}{15pt} \backslashbox{Cutoff}{$\left<z\right>$}  &  $0.33$ & $0.54$ & $0.74$ & $1.01$ & $1.62$ \\
\hline \hline
Cutoff 1 & 1.4 & 1.1 & 0.9 & 0.9 & 0.8 \\\hline
Cutoff 2 & 2.9 & 2.2 & 1.9 & 1.7 & 1.6 \\\hline
Cutoff 3 & 5.7 & 4.3 & 3.8 & 3.4 & 3.3 \\\hline
\end{tabular}
\caption{Approximate scales in wavenumber $k$, measured in $h$Mpc$^{-1}$, for each galaxy source bin that correspond to the angular cutoffs in $\xi_+$  tested in our LSST-like cosmic shear analysis. The scales are approximated by computing the angular diameter distance to the mean redshift of each bin and converting to a wavenumber.}
\label{tab:k_cuts}
\end{table}


\subsection{Quantifying discrepancies}
To quantify deviations between parameter constraints from the LSST-Y1 simulated analyses using the nonlinear prescriptions $X$ and $Y$, we first evaluate the one-dimensional bias for key parameters $\Omega_m$, $S_8$, $w_0$, and $w_a$, defined as
\begin{equation}
    \frac{\Delta \theta_{i}}{\sigma_{\theta_{i}}} = \frac{\braket{\theta_{i}}_\mathrm{X} - \braket{\theta_{i}}_\mathrm{Y}}{\sqrt{\sigma_{\theta_{i},\mathrm{X}}^2 + \sigma_{\theta_{i},\mathrm{Y}}^2}},
    \label{eq:1d_fob}
\end{equation}
where $\theta_{i}$ denotes one of $\Omega_m$, $S_8$, $w$, or $w_a$, and $\braket{\theta_{i}}$ and $\sigma^2_{\theta_{i}}$ are, respectively, the sample mean and sample variance from MCMC posteriors.

To capture parameter correlations, we also compute the Figure of Bias (FoB), a multivariate generalization of the 1D bias defined by 
\begin{equation}
    \mathrm{FoB}(\boldsymbol{\theta}) = [\Delta\!\braket{\boldsymbol{\theta}}^\mathrm{T} \cdot (C_\mathrm{COLA} + C_\eetwo)^{-1} \cdot \Delta\!\braket{\boldsymbol{\theta}}]^{1/2},
    \label{eq:fob}
\end{equation}
where $\boldsymbol{\theta}$ denotes a vector of cosmological parameters, $\Delta\!\braket{\boldsymbol{\theta}} = \braket{\boldsymbol{\theta}}_\mathrm{COLA} - \braket{\boldsymbol{\theta}}_\eetwo$ is the difference in sample means, and $C_\mathrm{X}$ denotes the parameter covariance matrices for prescription $\mathrm{X}$. We choose to calculate the FoB in selected 2D planes: $\Omega_m \times S_8$, $\Omega_m \times w_0$ and $\Omega_m \times w_a$. Furthermore, we also calculate the FoB in the seven cosmological parameters. A bias of less than $0.3$ is considered negligible \cite{des_y3_strategy}.

Changing the nonlinear modeling of the cosmic shear data vector may lead to underestimating or overestimating cosmological parameters, compared to a fiducial model. The strength of the constraints is measured by the Figure of Merit (FoM) statistic, defined as
\begin{equation}\label{eq:fom}
    \mathrm{FoM} = \alpha\det(C)^{-1/2},
\end{equation}
where $C$ is the covariance matrix of cosmological parameters obtained from the MCMC, and $\alpha$ is a prefactor that depends on the desired limits (\textit{i.e.}, $1\sigma$ or $2\sigma$) and the number of parameters considered \cite{albrecht2006reportdarkenergytask}. We report the FoM ratio between the COLA and benchmark analyses.

To assess whether a nonlinear prescription $\mathrm{X}$ may degrade the goodness-of-fit when compared to the benchmark emulator, we compute the quantity
\begin{equation}\label{eq:dchi2}
    \Delta\chi^2 = (\mathbf{t}_\mathrm{X} - \mathbf{t}_\mathrm{EE2})^T \cdot C_\mathrm{data}^{-1} \cdot (\mathbf{t}_\mathrm{X} - \mathbf{t}_\mathrm{EE2}),
\end{equation}
where $\mathbf{t}_\mathrm{X}$ is the cosmic shear theory prediction calculated using the nonlinear prescription $\mathrm{X}$ and $C_\mathrm{data}$ is the data covariance matrix. We compute this quantity for random points across the parameter space.

\section{Results for LSST-Y1 Simulated Analysis}
\label{sec:Results}
\begin{figure*}[ht] 
    \centering
    \includegraphics[width=0.33\linewidth]{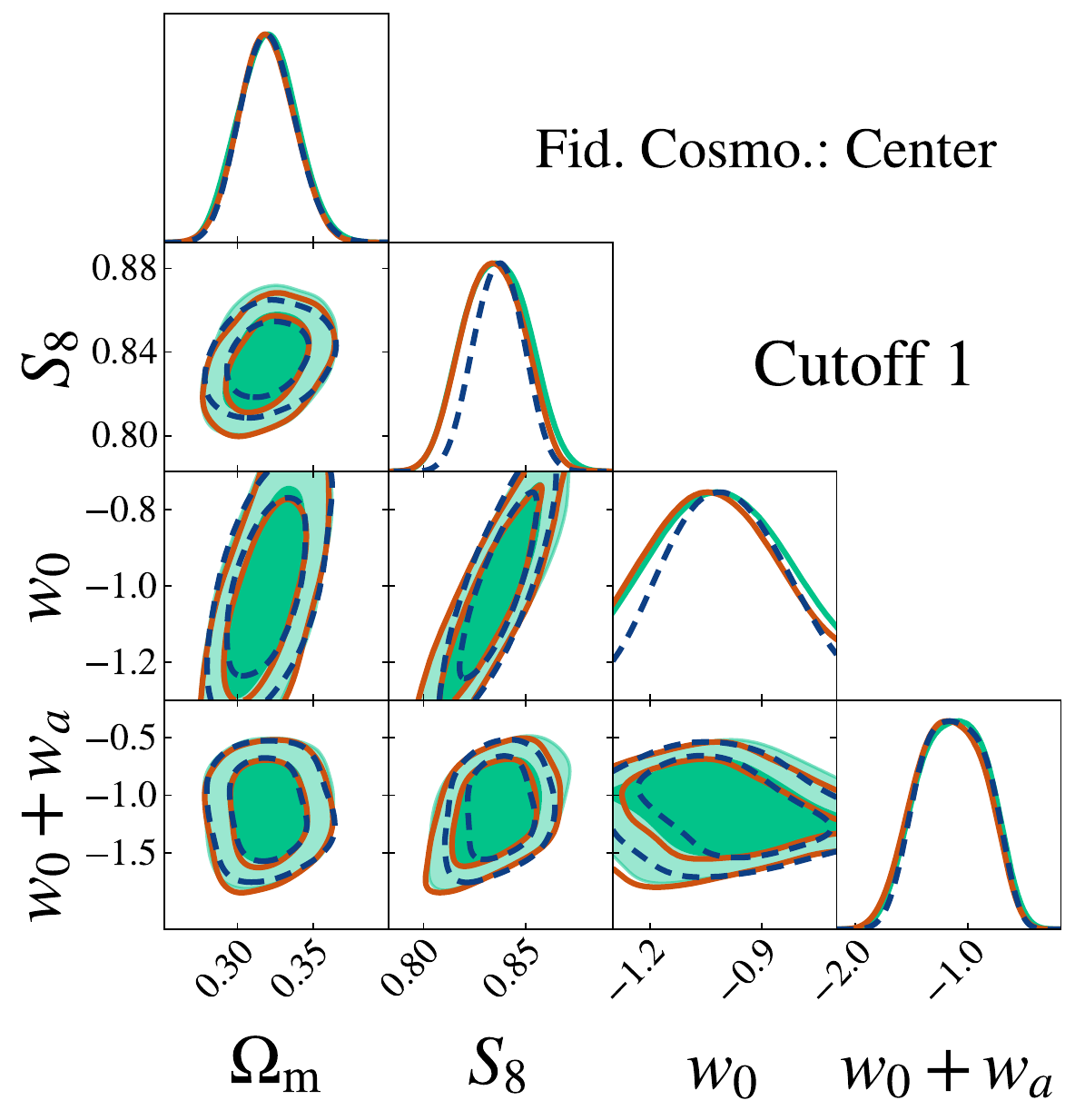}
    \includegraphics[width=0.33\linewidth]{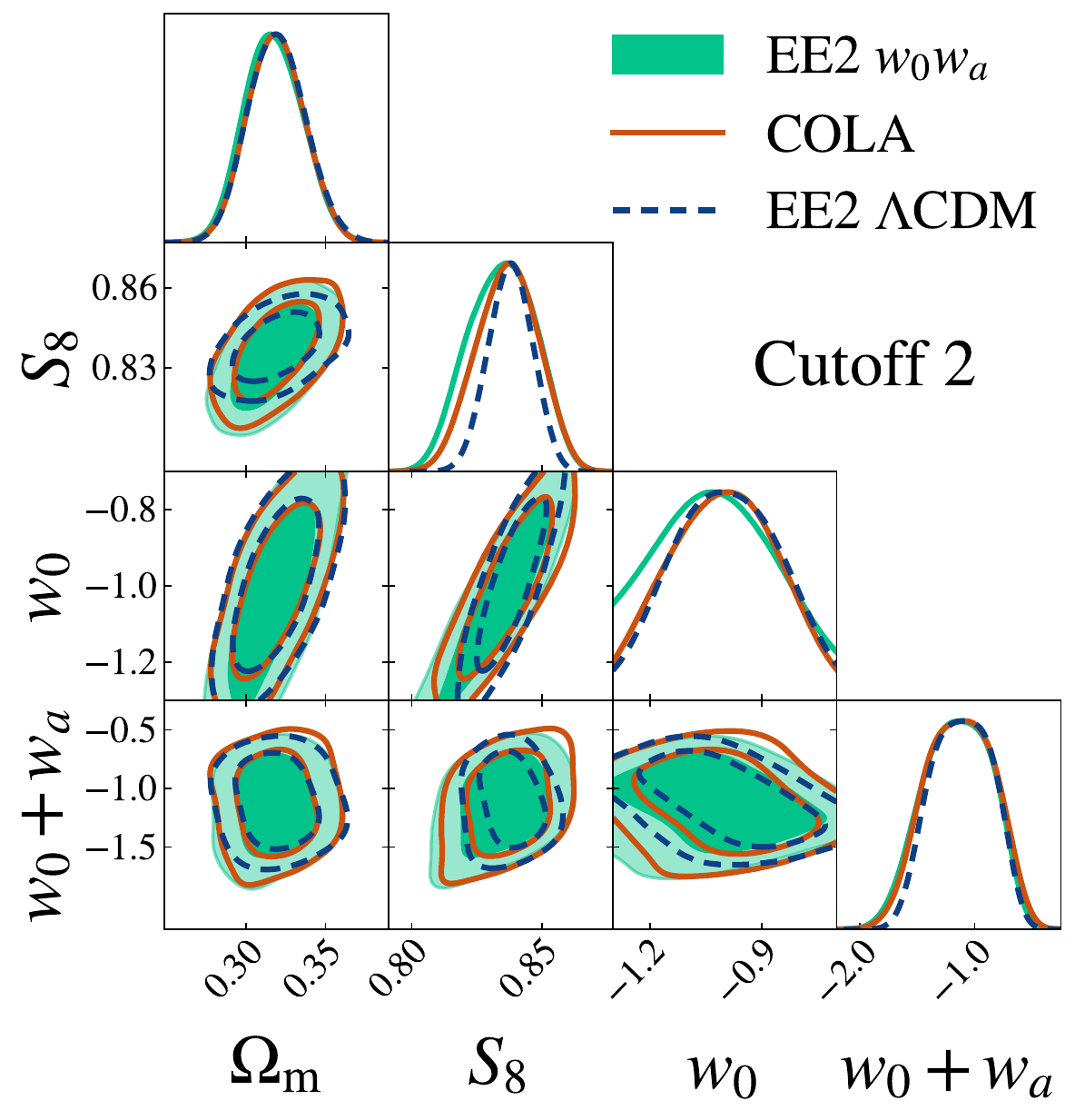}
    \includegraphics[width=0.33\linewidth]{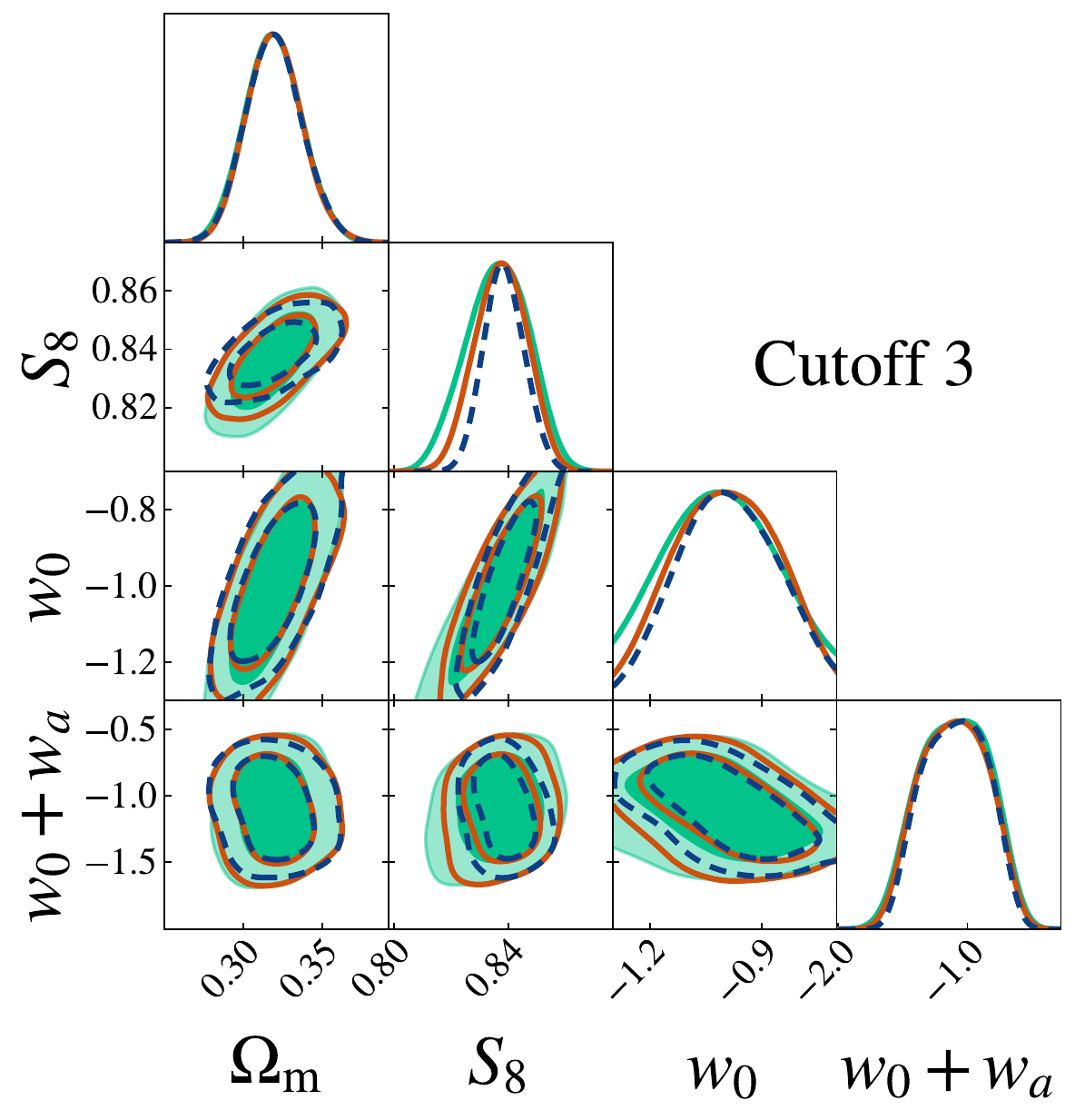}
    \caption{Cosmological parameter constraints (68\% and 95\%) from the LSST-Y1 simulated analyses assuming the center cosmology from Table~\ref{tab:param_space} as the fiducial. Green-filled contours denote constraints obtained using \euclidemutwo~as the nonlinear prescription, orange dashed and dotted contours use our COLA emulator, and blue dashed contours use \eetwo~$\Lambda$CDM prescription. The left, middle, and right panels show constraints using the angular cutoffs C1, C2, and C3, respectively. We observe no shifts between the analyses; however, using cutoffs C2 and C3, the constraints obtained using COLA are slightly tighter than those found with \eetwo~for $S_8$ and $w_0$. For \eetwo~$\Lambda\mathrm{CDM}$, this effect is amplified.}
    \label{fig:triangle-ee2ref}
\end{figure*}

To evaluate the accuracy of our emulator in practice, we begin by examining cosmological parameter constraints across three nonlinear prescriptions: the \euclidemutwo~benchmark, our COLA-based emulator, and \eetwo~$\Lambda$CDM (\textit{i.e.}, using the boost from the projected $\Lambda$CDM cosmology). Figure~\ref{fig:triangle-ee2ref} shows 1D and 2D posterior contours (68\% and 95\%) for the parameters $\Omega_m$, $S_8$, $w_0$ and the sum $w_0+w_a$,  assuming the central cosmology (see Table~\ref{tab:param_space}) as the fiducial. Since the constraints on $w_a$ are weak, we focus on $w_0+w_a$ as the more informative parameter combination. 

Because our fiducial cosmology is in the $\Lambda$CDM subspace, where all prescriptions are equivalent, we observe no significant biases between the different constraints, and all posteriors peak at the same point. However, an indication of failure outside $\Lambda$CDM is observed in the size of the error bars. Employing the most conservative cutoff (C1), the COLA and \euclidemutwo~contours are nearly indistinguishable. At the same time, those of \eetwo~$\Lambda$CDM are significantly smaller than the "true" error bars of \eetwo, especially at the parameters $S_8$ and $w_0$, positively correlated. These results suggest that, in the context of LSST-Y1 cosmic shear, COLA emulators are equivalent to high-precision \textit{N}-body emulators up to $k = 1 \, h/\mathrm{Mpc}$.

\edit{The middle panel of Figure~\ref{fig:errors} indicates that, as we advance towards smaller scales}, the disagreements in the boost predictions between COLA and \euclidemutwo~quickly increase. For Cutoffs 2 and 3, the constraints are still in excellent agreement, but COLA yields slightly tighter error bars on $S_8$ and $w_0$; this effect is much more pronounced in the case of \eetwo~$\Lambda$CDM. The 1D marginalized constraints under Cutoff 2 are:
\begin{itemize} 
    \item \euclidemutwo: $S_8 = 0.835\pm 0.013$, $w_0 = -1.02^{+0.15}_{-0.18}$, $w_0 + w_a = -1.15^{+0.32}_{-0.28}$;
    \item COLA: $S_8 = 0.837\pm 0.012$, $w_0 = -1.00\pm 0.14$, $w_0 + w_a = -1.13\pm 0.29$;
    \item \eetwo~$\Lambda$CDM: $S_8 = 0.838\pm 0.008$, $w_0 = -1.00\pm 0.14$, $w_0 + w_a = -1.11\pm 0.26$.
\end{itemize}
Using Cutoff 3, the constraints are:
\begin{itemize} 
    \item \euclidemutwo: $S_8 = 0.836^{+0.012}_{-0.010}$, $w_0 = -1.01 \pm 0.14$, $w_0 + w_a = -1.10 \pm 0.26$;
    \item COLA: $S_8 = 0.838 \pm 0.009$, $w_0 = -1.00 \pm 0.13$, $w_0 + w_a = -1.10 \pm 0.26$;
    \item \eetwo~$\Lambda$CDM: $S_8 = 0.838\pm 0.007$, $w_0 = -0.99\pm 0.13$, $w_0 + w_a = -1.09\pm 0.24$.
\end{itemize}

To quantify the overestimation in $S_8$ and $w_0$, we compute the figure of merit (FoM) in the $S_8 \times w_0$ plane. Assuming Cutoff 2, relative to \euclidemutwo, the COLA emulator increases the FoM by 8\%, indicating slightly tighter constraints, whereas the projected \eetwo~$\Lambda$CDM boost inflates the FoM by approximately 47\%. For the most "aggressive" Cutoff 3, the FoM obtained with COLA is 19\% bigger than that of \eetwo, while \eetwo~$\Lambda$CDM increases the FoM by 58\%. Remarkably, in terms of figure of merit, the COLA emulator performs better at Cutoff 3 than \eetwo~$\Lambda$CDM at Cutoff 1, where the FoM is increased by 28\%. From Figure~\ref{fig:triangle-ee2ref}, we observe that most of the disagreement between COLA and \euclidemutwo~ lies in low $w_0$ and low $S_8$ values, two parameters that are positively correlated in the analysis. This region of the parameter space is excluded by low-redshift geometric data from type-Ia supernovae~\cite{pantheonplus, desy5} and BAO~\cite{desi_dr1, desi_dr2_bao}. Therefore, we expect that this disagreement would not affect constraints obtained from the combination of LSST cosmic shear data with supernovae and BAO distance measurements.

\begin{figure} 
    \centering
    \includegraphics[width=0.98\linewidth]{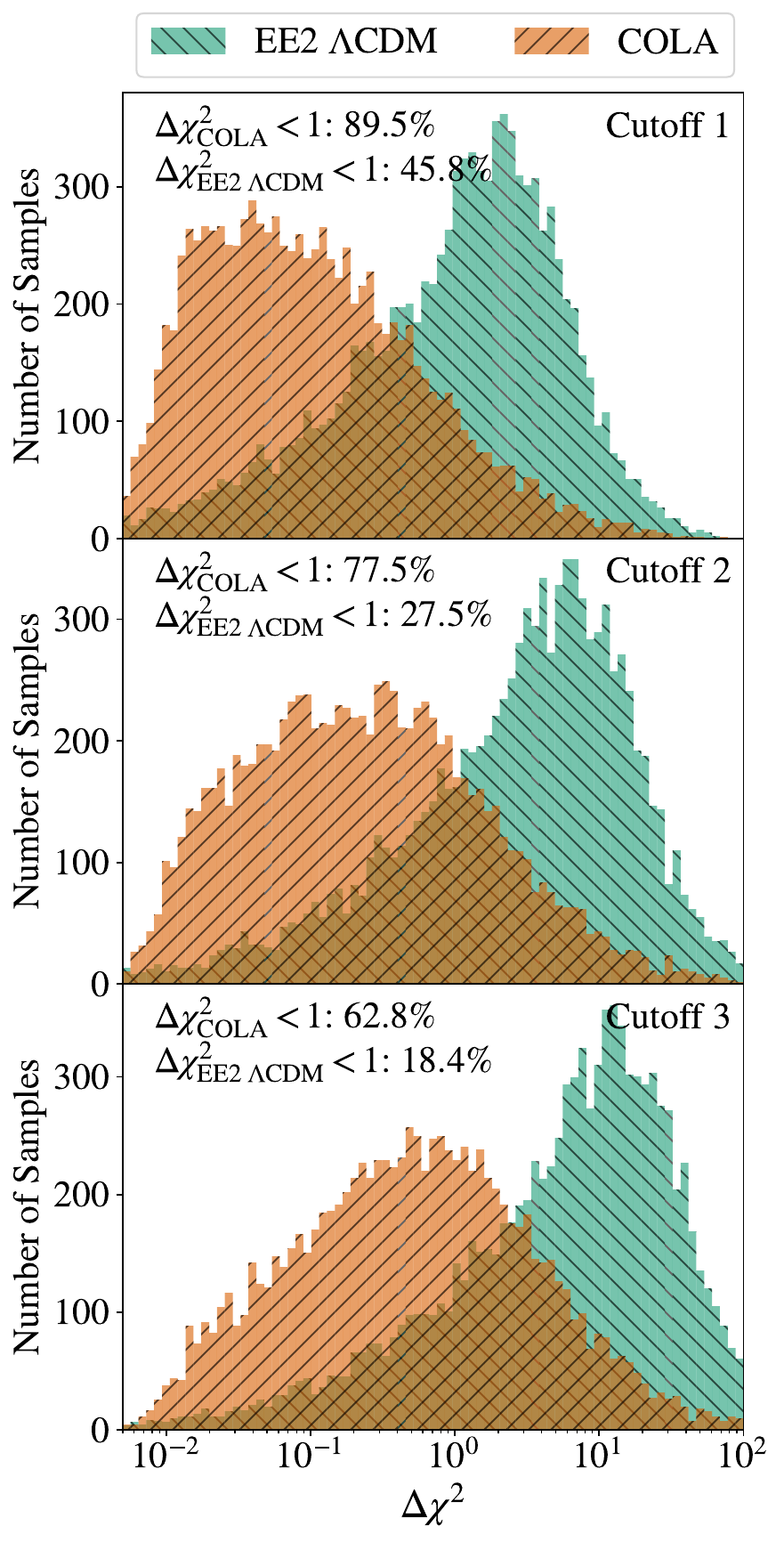}
    \caption{Histograms of $\Delta\chi^2_\mathrm{X} = (\mathbf{t}_\mathrm{X} - \mathbf{t}_\mathrm{EE2})^T \cdot C_\mathrm{data}^{-1} \cdot (\mathbf{t}_\mathrm{X} - \mathbf{t}_\mathrm{EE2})$ for prescription $\mathrm{X} \in \{\mathrm{COLA},\mathrm{EE2}\;\Lambda \mathrm{CDM}\}$ compared to \eetwo~at the full $w_0w_a$ cosmology, with random samples drawn from the prior. The top, middle, and bottom panels show results for angular cutoffs C1, C2, and C3, respectively. The distribution of $\Delta\chi^2$ values demonstrates an order of magnitude difference between theory predictions calculated using our COLA method compared to the traditional \eetwo~ $\Lambda$CDM approach, showing an improved fit from modeling the extended parameters using COLA.}
    \label{fig:dchi2-histogram}
\end{figure}

\begin{figure}
    \centering
    \includegraphics[width=\linewidth]{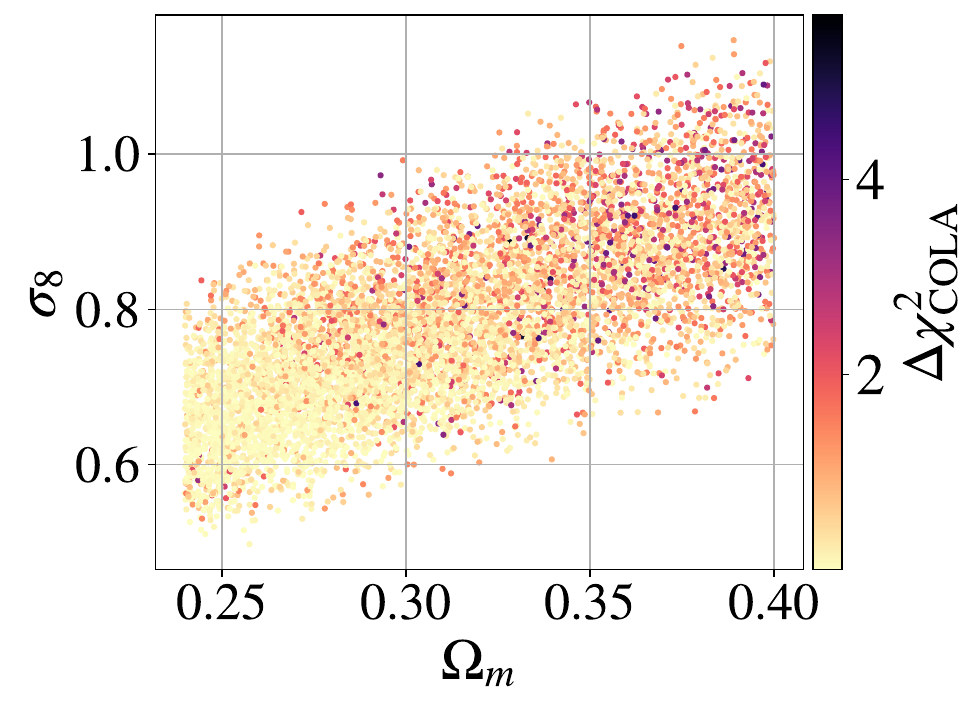}
    \caption{Spatial distribution of $\Delta\chi^2$ in the $\Omega_m \times \sigma_8$ plane. We observe higher values of $\Delta\chi^2$ at higher values of $\Omega_m$ and $\sigma_8$.}
    \label{fig:dchi2-omegam-sigma8}
\end{figure}

Figure~\ref{fig:dchi2-histogram} further investigates the disagreements between COLA and \euclidemutwo, showing histograms of $\Delta\chi^2$ (see Equation~\ref{eq:dchi2}) for COLA and \eetwo~$\Lambda$CDM compared to \euclidemutwo, obtained from 10,000 cosmologies sampled randomly from the emulation box (see Table~\ref{tab:param_space}). Across all cutoffs, the \eetwo~$\Lambda$CDM prescription yields $\Delta\chi^2$ distributions that are substantially broader than those from our COLA emulator. As such, the use of COLA simulations can significantly improve the consistency with the high-precision benchmark, while the projection approach fails to capture the nonlinear corrections required by dynamical dark energy models, leading to degraded fits and potentially biased constraints. In Figure~\ref{fig:dchi2-omegam-sigma8}, we investigate how these $\Delta\chi^2$ values distribute around cosmological space; we find a clear trend of higher values of $\Delta\chi^2$ correlated with higher values of $\Omega_m$ and $\sigma_8$.

\begin{figure*}[ht] 
    \centering
    \includegraphics[width=0.33\linewidth]{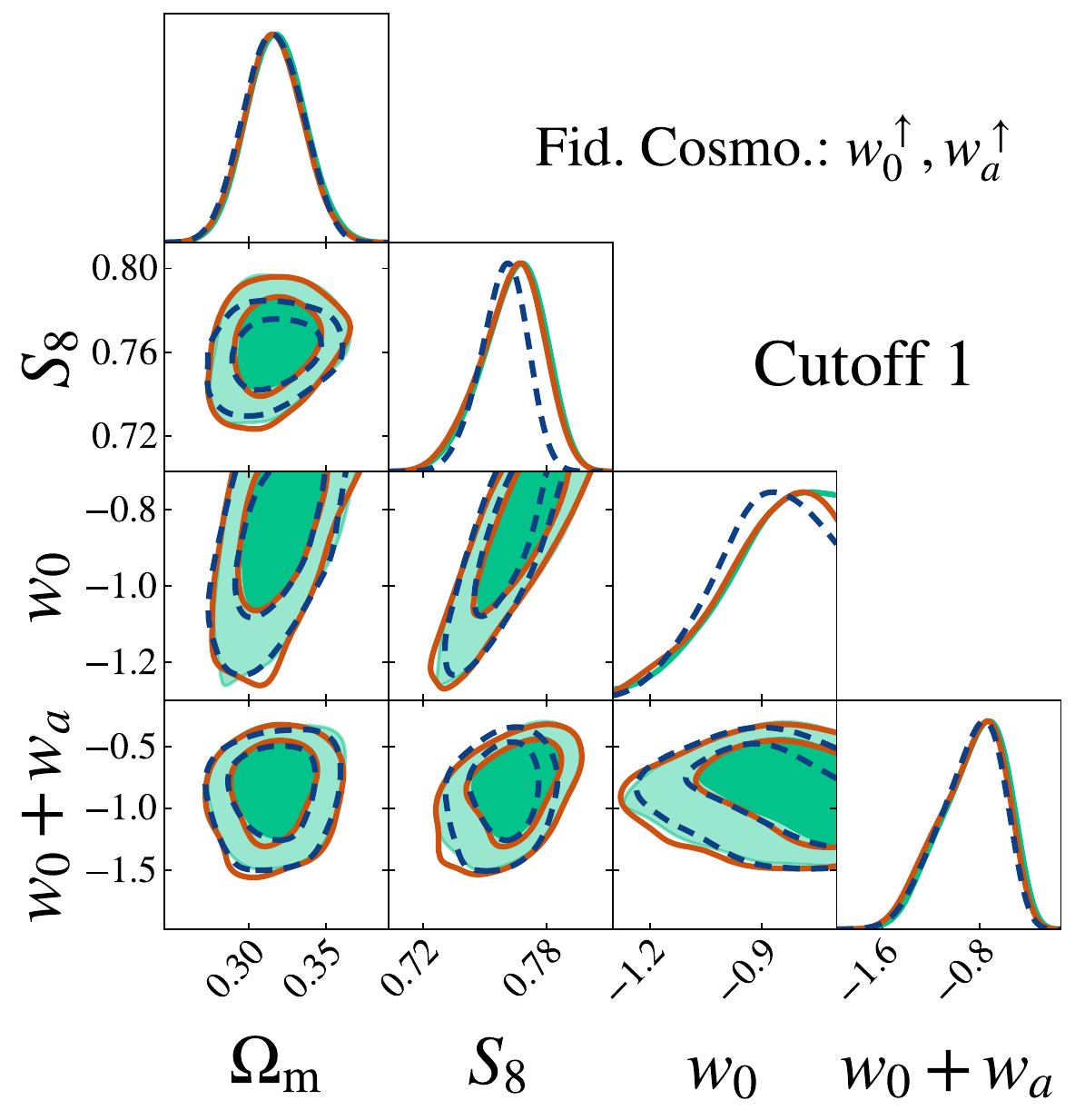}
    \includegraphics[width=0.33\linewidth]{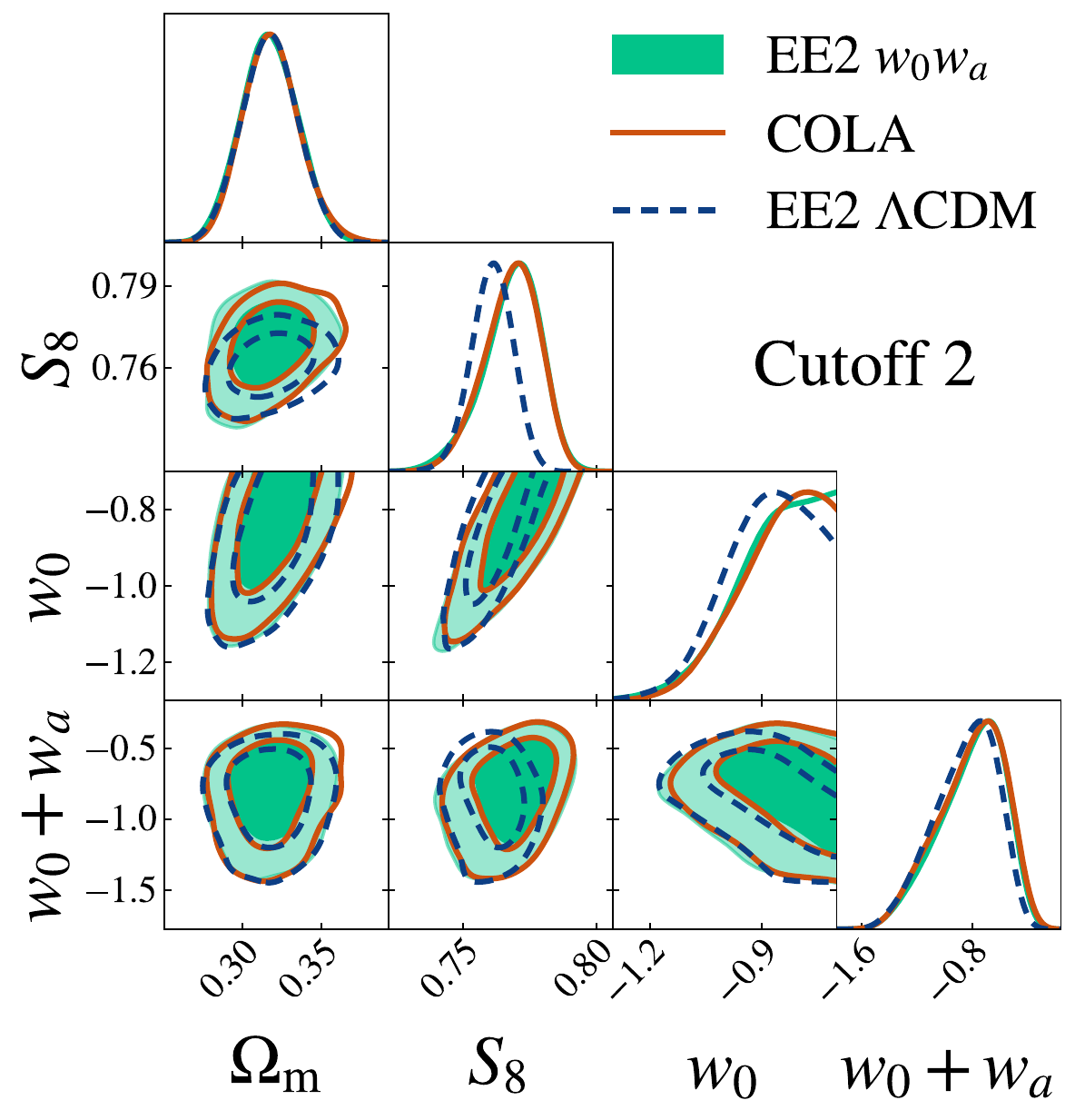}
    \includegraphics[width=0.33\linewidth]{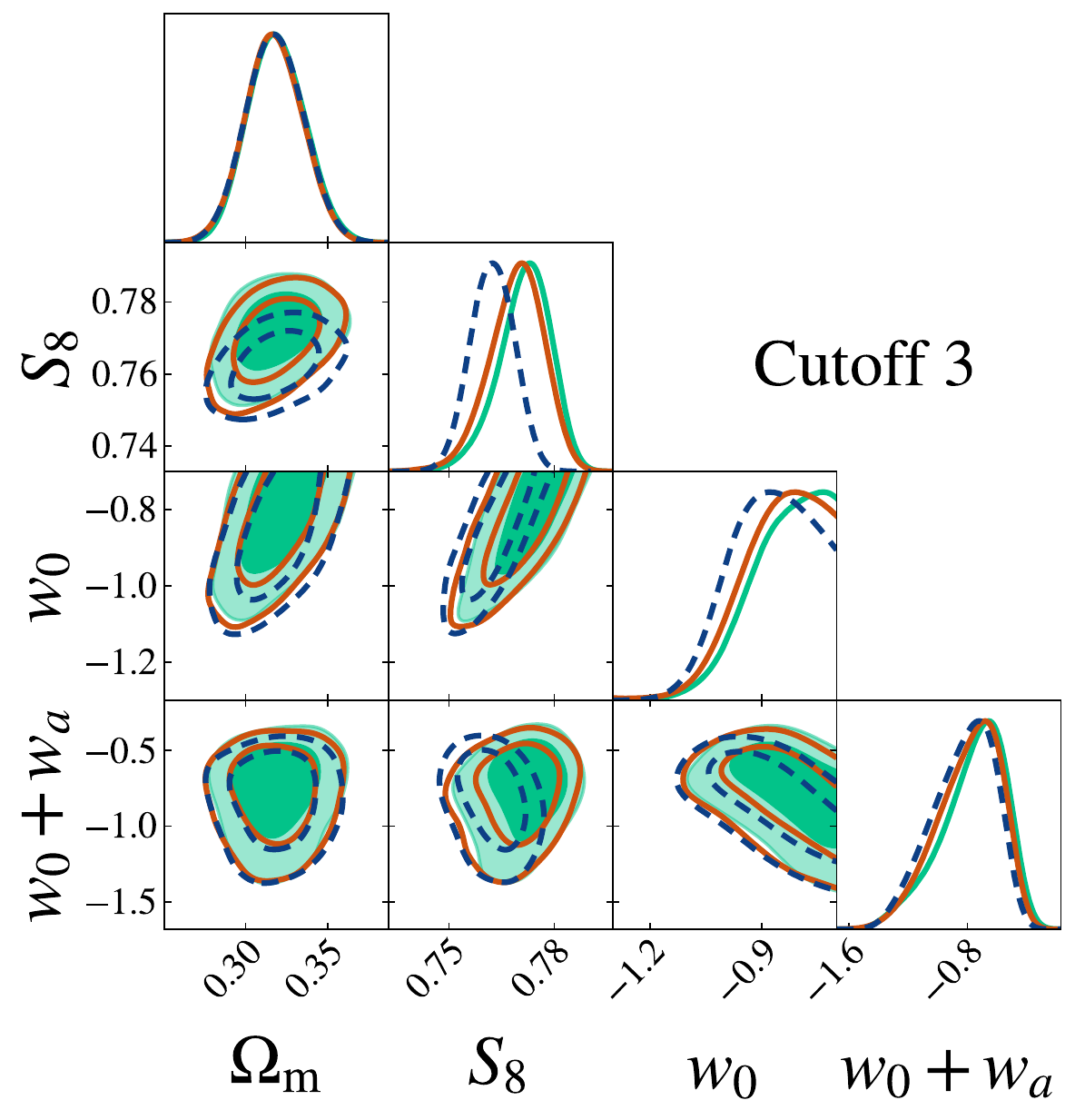}
    \caption{Cosmological parameter constraints (68\% and 95\%) from the LSST-Y1 simulated analyses assuming a fiducial cosmology with $w_0^\uparrow$ and $w_a^\uparrow$ (see Table~\ref{tab:fiducials}), keeping the other parameters at their central values. Green-filled contours denote constraints obtained using \euclidemutwo~as the nonlinear prescription, orange dashed and dotted contours use our COLA emulator, and blue dashed contours use \eetwo~$\Lambda$CDM prescription. The left, middle, and right panels show constraints using the angular cutoffs C1, C2, and C3, respectively. In this case, the \eetwo~$\Lambda$CDM prescription can provide significant biases in $S_8$, which are not present when using COLA.}
    \label{fig:triangle-whi_wahi}
\end{figure*}

To assess the robustness of our results beyond the $\Lambda$CDM subspace, we repeat our analysis using the fiducials described in Section~\ref{sec:Methodology}, all of them with $w_0 \neq -1$ and $w_a \neq 0$. We find that the biases may increase as we shift $w_0$ and $w_a$ in the same direction, either higher or lower, in the parameter space of Table~\ref{tab:param_space}. This is due to a known geometrical degeneracy along $w_0 + w_a$, which can suppress modeling systematics when $w_0 + w_a \approx 1$. Figure~\ref{fig:triangle-whi_wahi} illustrates the results for a fiducial with $w_0^\uparrow$ and $w_a^\uparrow$ (see Table~\ref{tab:fiducials}), keeping the other parameters fixed to their central values. In this case, we see a remarkable agreement between COLA and \eetwo~ across all angular cutoffs. We focus on Cutoff 3, which yields the marginalized 1D constraints:
\begin{itemize} 
    \item \euclidemutwo: $S_8 = 0.772^{+0.008}_{-0.006}$, $w_0 > -0.882$, $w_0 + w_a = -0.76^{+0.26}_{-0.15}$;
    \item COLA: $S_8 = 0.770^{+0.008}_{-0.007}$, $w_0 > -0.900$, $w_0 + w_a = -0.79^{+0.26}_{-0.17}$;
    \item \eetwo~$\Lambda$CDM: $S_8 = 0.762\pm 0.006$, $w_0 = -0.879^{+0.140}_{-0.088}$, $w_0 + w_a = -0.82^{+0.26}_{-0.17}$.
\end{itemize}
In this case, the \eetwo~$\Lambda\mathrm{CDM}$ projection induces substantial biases in $S_8$, shifting up by nearly 1$\sigma$. By comparison, the COLA emulator remains consistent to within 0.25 standard deviations. This trend extends to the multidimensional figures of bias: COLA has a 7D figure of bias of $\mathrm{FoB}_\mathrm{7D} = 0.27$ compared to the benchmark, while the projected \eetwo~$\Lambda$CDM reaches $\mathrm{FoB}_\mathrm{7D} = 1.04$.

Unlike the parameter constraints assuming the center cosmology from Table \ref{tab:param_space} as the fiducial, we see in Figure \ref{fig:triangle-whi_wahi} that the posteriors calculated using our COLA-based emulator now closely track those generated using our $N$-body proxy, rather than overestimating $S_8$ and $w_0$ in any substantive way. We find that in switching from the center cosmology to $w_0^\uparrow$ and $w_a^\uparrow$, the relative ratio between FoMs of COLA and \eetwo~in the $S_8\times w_0$ plane is $0.97$, while the same ratio is $1.21$ for \eetwo~$\Lambda$CDM.

\begin{figure*} 
    \centering
    \includegraphics[width=0.9\linewidth]{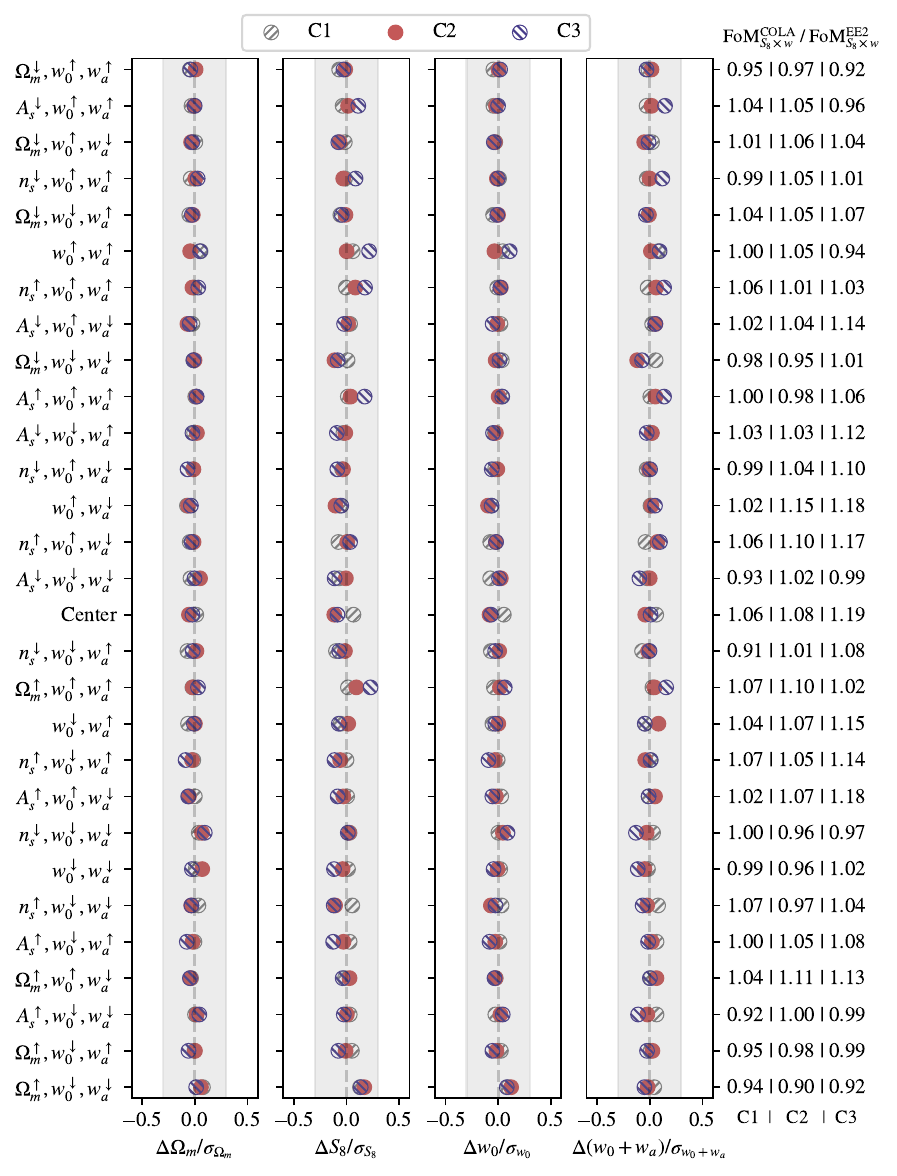}
    \caption{One-dimensional biases from Equation~\ref{eq:1d_fob}, shown for the parameters $\Omega_m$, $S_8$, $w_0$ and $w_a$, sorted in increasing order of their associated $\sigma_8$ values. On the right-hand side, we show the ratios in figure of merit (see Equation~\ref{eq:fom}) in the $ S_8\times w_0$ plane between the analyses using COLA and \eetwo, for the three angular cutoffs. The gray bands represent $0.3\sigma$ bias.}
    \label{fig:1d_biases}
\end{figure*}

To investigate whether our COLA emulator can provide unbiased constraints with FOMs similar to \eetwo~across the parameter space, Figure \ref{fig:1d_biases} shows the 1D biases of Equation~\ref{eq:1d_fob} for $\Omega_m$, $S_8$, $w_0$ and $w_a$ between the COLA emulator and our baseline \eetwo~for all scale cuts and all of the 29 cosmologies outlined in Section \ref{sec:Methodology}. The fiducial cosmologies are listed in increasing order of their associated $\sigma_8$ values. All 1D biases are within $0.3\sigma$, even for cosmologies with higher values of $\sigma_8$ and $\Omega_m$, where Figure~\ref{fig:dchi2-omegam-sigma8} shows our emulator performs worse. We computed the 7D figure of bias in cosmological parameters for all fiducial cosmologies using Cutoff 3, finding a maximum value of $\mathrm{FoB}_\mathrm{7D} = 0.35$ at the cosmology $\Omega_{m}^\uparrow$, $w_0^\uparrow$, $w_a^\uparrow$. We conclude that, for all scale cuts considered in this work, our emulator succeeds in providing unbiased constraints on cosmological parameters when compared to a high-precision \textit{N}-body emulator in the context of dynamical dark energy models, even with significant variations in cosmological parameters and extreme values of $\sigma_8$. Further, Figure~\ref{fig:1d_biases} indicates our measure of the relative tightness of the parameter constraints in the $S_8\times w$ plane compared to \eetwo, FOM$^\mathrm{COLA}_{S_8\times w}$/FOM$^\mathrm{EE2}_{S_8\times w}$, assuming Cutoff 3. The highest ratio is $1.19$ for the center cosmology, shown in Figure~\ref{fig:triangle-ee2ref}. The same remark from before applies to other fiducial cosmologies: the disagreement between COLA and \eetwo~is driven mainly by low values of $w_0$ and $S_8$, a region of the parameter space disallowed by low-redshift distance measurements.

\section{Conclusion}
\label{sec:Conclusion}
In light of recent hints of dynamical dark energy, constraining its equation of state has become a task of central importance. While geometrical probes, such as BAO and supernovae measurements, are the primary probes of late dark energy behavior, galaxy surveys can also probe dark energy dynamics through their effects on the growth of large-scale structure. Extracting robust constraints from these surveys requires accurate modeling of nonlinear gravitational effects in the matter power spectrum, which becomes increasingly challenging in extended cosmologies where $w(z)$ departs from a cosmological constant. While accurate modeling can be achieved with $N$-body simulations, their high computational cost limits their applicability across the myriad candidate dynamical dark energy models (\textit{e.g.}, quintessence), and, further, for models that directly impact the growth of matter perturbations beyond modifications in the Universe's expansion. In this context, COLA is a fast approximate alternative to $N$-body simulations, which, when appropriately corrected, presents an avenue for constructing accurate emulators for the nonlinear matter power spectrum at a fraction of the computational cost. 

In this work, we have built an emulator for the nonlinear boost assuming the $w_0w_a$CDM cosmological model using a suite of 1400 COLA simulations, two for each of the 700 cosmologies in the training set to account for pairing-and-fixing. To evaluate the accuracy of the neural network, we ran a pair of simulations for each of the 200 cosmologies in the test set. The total computational cost of all simulations is estimated at 153.600 CPU-hours. A simple connected neural network with a trainable activation function can reproduce the test set boosts at $0.1\%$ error. The computational cost of the simulation suite could potentially be lowered by using an alternative sampling algorithm for the training set cosmologies: the Sobol sequence~\cite{Sobol}, which has been shown to improve emulation errors compared to Latin hypercube sampling~\cite{Chen_2025}.

We have compared our COLA emulator to a benchmark $N$-body emulator, chosen as \euclidemutwo. We test an additional nonlinear prescription common to analyses of extended cosmological models without $N$-body simulations: using $N$-body boosts at the projected $\Lambda$CDM cosmology (\textit{i.e.}, setting $w_0 = -1$ and $w_a$ = 0), an approach we denote as \eetwo~$\Lambda$CDM. We compare nonlinear models in two manners: at the boost level, shown in Figure~\ref{fig:errors}, and at the level of a simulated cosmic shear analysis akin to LSST-Y1, assuming \euclidemutwo~ as the true nonlinear model. In the data analysis, to account for possible variations in the cosmological parameters when beyond-$\Lambda$CDM models are analyzed, we define fiducial cosmologies scattered across the parameter space, including outside $\Lambda$CDM, with significant variations in $\Omega_m$ and $\sigma_8$. We have assessed the goodness-of-fit degradation, shown in Figure~\ref{fig:dchi2-histogram}, and parameter constraint biases, shown in Figures \ref{fig:triangle-ee2ref}, \ref{fig:triangle-whi_wahi} and \ref{fig:1d_biases}. We find that, at the boost level, our emulator can reproduce the benchmark emulator results with less than $2\%$ error at $k = 1 \, h/\mathrm{Mpc}$, while \textsc{EE2} $\Lambda$CDM produces $7.5\%$ errors at the same scales. As for the simulated analysis, we find that our COLA-based emulator can provide unbiased constraints compared to \euclidemutwo: all 1D biases are well within $0.3\sigma$, even for the most aggressive angular cutoffs and exotic fiducial cosmologies with extreme values of $\Omega_m$, $\sigma_8$, or outside $\Lambda$CDM. Furthermore, all 7D figures of bias are below 0.35. At the precision level expected for the first year of LSST observations, the COLA emulator yields constraints equivalent to those obtained using \euclidemutwo~for scales up to $k \approx 3 \, h/\mathrm{Mpc}$, comparable to our Cutoff 2. \edit{We used cosmic shear as a simple test, but our methodology also applies to the 3x2pt analysis, which includes galaxy clustering and galaxy-galaxy lensing 2-point correlation functions, as well as CMB lensing.}

\edit{At the scales investigated in our work, the matter power spectrum is affected not only by nonlinear gravitational effects, but also baryonic feedback processes~\cite{Mead:2015yca, Mead:2016zqy, Mead:2020vgs}, which suppress the matter power spectrum at very small scales. Studies about correcting COLA simulations with baryonic effects~\cite{cola_baryons, baryonification} are left for future work. }

Our results demonstrate that COLA, when combined with an accurate $\Lambda$CDM reference, offers a viable and flexible framework for extending nonlinear modeling to dynamical dark energy and other beyond-$\Lambda$CDM scenarios. We emphasize that, while we use \euclidemutwo~ as the "baseline" $\Lambda$CDM emulator in Equation~\ref{eq:inf_refs}, any other emulator could be used to provide $\Lambda$CDM boosts. Moreover, our methodology can be applied to more exotic models that also modify the growth of structure directly, such as modified gravity or coupled dark energy~\cite{ide_1}. We also remark that there are avenues to improve our methodology. One example is the choice of "reference" cosmology. Equation~\ref{eq:inf_refs} uses the projected $\Lambda$CDM cosmology because, geometrically, it is the closest cosmology; however, a possible choice that improves accuracy is to use a $w$CDM cosmology with the same value of $\sigma_8$, akin to what is done in the Casarini~\cite{casarini} prescription. Moreover, COLA can be combined with other analytical or semi-analytical prescriptions, such as the one proposed in~\cite{web_halo_model}, improving its accuracy at small scales.

\section*{Acknowledgements}
The authors would like to thank Stony Brook Research Computing and Cyberinfrastructure and the Institute for Advanced Computational Science at Stony Brook University for access to the high-performance SeaWulf computing system. This research was also supported by resources supplied by the Center for Scientific Computing (NCC/GridUNESP) of the São Paulo State University (UNESP). This work made use of the CHE cluster, managed and funded by COSMO/CBPF/MCTI, with financial support from FINEP and FAPERJ, and operating at the RSDC - Datacenter para Ciência Alfredo Marques de Oliveira/CBPF. JR acknowledges the financial support from FAPESP under grant 2020/03756-2, São Paulo Research Foundation (FAPESP) through ICTP-SAIFR. JR also acknowledges that this study was financed in part by the Coordenação de Aperfeiçoamento de Pessoal de Nível Superior – Brasil (CAPES) – Finance Code 001. VM is supported by the Roman Project Infrastructure Team ``Maximizing Cosmological Science with the Roman High Latitude Imaging Survey" (NASA contracts 80NM0018D0004-80NM0024F0012). V.M. is also partially supported by the Roman Project Infrastructure Team ``A Roman Project Infrastructure Team to Support Cosmological Measurements with Type Ia Supernovae" (NASA contract 80NSSC24M0023).

\section*{Data Availability}
The LSST-Y1 simulated cosmic shear dataset was generated using \textsc{Cocoa}~\cite{cocoa}. The COLA simulations are generated using \textsc{COLA-FML}~\cite{fml}. The training and test datasets, the emulator codes and other data products, cannot be made publicly available because it is not technically feasible and/or the cost of preparing, depositing, and hosting the data would be prohibitive within the terms of this research project. The data are available from the authors upon reasonable request.

\bibliography{nls_short}

@INPROCEEDINGS{spherex,
       author = {{Crill}, Brendan P. and {Werner}, Michael and {Akeson}, Rachel and {Ashby}, Matthew and {Bleem}, Lindsey and others},
        title = "{SPHEREx: NASA's near-infrared spectrophotometric all-sky survey}",
    booktitle = {Space Telescopes and Instrumentation 2020: Optical, Infrared, and Millimeter Wave},
         year = 2020,
       editor = {{Lystrup}, Makenzie and {Perrin}, Marshall D.},
       series = {Society of Photo-Optical Instrumentation Engineers (SPIE) Conference Series},
       volume = {11443},
        month = dec,
          eid = {114430I},
        pages = {114430I},
          doi = {10.1117/12.2567224},
archivePrefix = {arXiv},
       eprint = {2404.11017},
 primaryClass = {astro-ph.IM},
       adsurl = {https://ui.adsabs.harvard.edu/abs/2020SPIE11443E..0IC}
}

@ARTICLE{cosmology_intertwined_s8,
      author = {{Di Valentino}, Eleonora and {Anchordoqui}, Luis A. and {Akarsu}, {\"O}zg{\"u}r and {Ali-Haimoud}, Yacine and {Amendola}, Luca and others},
        title = "{Cosmology Intertwined III: f{\ensuremath{\sigma}}$_{8}$ and S$_{8}$}",
      journal = {Astroparticle Physics},
         year = 2021,
        month = sep,
       volume = {131},
          eid = {102604},
        pages = {102604},
          doi = {10.1016/j.astropartphys.2021.102604},
archivePrefix = {arXiv},
       eprint = {2008.11285},
 primaryClass = {astro-ph.CO},
       adsurl = {https://ui.adsabs.harvard.edu/abs/2021APh...13102604D}
}

@ARTICLE{roman_multiprobe,
      author = {{Eifler}, Tim and {Miyatake}, Hironao and {Krause}, Elisabeth and {Heinrich}, Chen and {Miranda}, Vivian and others},
        title = "{Cosmology with the Roman Space Telescope - multiprobe strategies}",
      journal = {\mnras},
         year = 2021,
        month = oct,
       volume = {507},
       number = {2},
        pages = {1746-1761},
          doi = {10.1093/mnras/stab1762},
archivePrefix = {arXiv},
       eprint = {2004.05271},
 primaryClass = {astro-ph.CO},
       adsurl = {https://ui.adsabs.harvard.edu/abs/2021MNRAS.507.1746E}
}

@ARTICLE{euclid_overview,
       author = {{Euclid Collaboration}},
        title = "{Euclid. I. Overview of the Euclid mission}",
      journal = {arXiv e-prints},
         year = 2024,
        month = may,
          eid = {arXiv:2405.13491},
        pages = {arXiv:2405.13491},
          doi = {10.48550/arXiv.2405.13491},
archivePrefix = {arXiv},
       eprint = {2405.13491},
 primaryClass = {astro-ph.CO},
       adsurl = {https://ui.adsabs.harvard.edu/abs/2024arXiv240513491E}
}

@article{Krause:2016jvl,
    author = "Krause, Elisabeth and Eifler, Tim",
    title = "{cosmolike \textendash{} cosmological likelihood analyses for photometric galaxy surveys}",
    eprint = "1601.05779",
    archivePrefix = "arXiv",
    primaryClass = "astro-ph.CO",
    doi = "10.1093/mnras/stx1261",
    journal = "Mon. Not. Roy. Astron. Soc.",
    volume = "470",
    number = "2",
    pages = "2100--2112",
    year = "2017"
}

@article{Lewis:1999bs,
      author         = "Lewis, Antony and Challinor, Anthony and Lasenby, Anthony",
      title          = "{Efficient computation of CMB anisotropies in closed FRW
                        models}",
      journal        = "\apj",
      volume         = "538",
      year           = "2000",
      pages          = "473-476",
      doi            = "10.1086/309179",
      eprint         = "astro-ph/9911177",
      archivePrefix  = "arXiv",
      primaryClass   = "astro-ph",
      SLACcitation   = "%%CITATION = ASTRO-PH/9911177;%%"
}

@article{Howlett:2012mh,
      author         = "Howlett, Cullan and Lewis, Antony and Hall, Alex and Challinor, Anthony",
      title          = "{CMB power spectrum parameter degeneracies in the era of
                        precision cosmology}",
      journal        = "Journal of Cosmology and Astroparticle Physics",
      volume         = "1204",
      year           = "2012",
      pages          = "027",
      doi            = "10.1088/1475-7516/2012/04/027",
      eprint         = "1201.3654",
      archivePrefix  = "arXiv",
      primaryClass   = "astro-ph.CO",
      SLACcitation   = "%%CITATION = ARXIV:1201.3654;%%"
}

@article{Lewis:2013hha,
    author = "Lewis, Antony",
    title = "{Efficient sampling of fast and slow cosmological parameters}",
    eprint = "1304.4473",
    archivePrefix = "arXiv",
    primaryClass = "astro-ph.CO",
    doi = "10.1103/PhysRevD.87.103529",
    journal = "Phys. Rev. D",
    volume = "87",
    number = "10",
    pages = "103529",
    year = "2013"
}

@article{Torrado:2020dgo,
    author = "Torrado, Jesus and Lewis, Antony",
    title = "{Cobaya: Code for Bayesian Analysis of hierarchical physical models}",
    eprint = "2005.05290",
    archivePrefix = "arXiv",
    primaryClass = "astro-ph.IM",
    reportNumber = "TTK-20-15",
    doi = "10.1088/1475-7516/2021/05/057",
    journal = "JCAP",
    volume = "05",
    pages = "057",
    year = "2021"
}

@ARTICLE{speculator,
      author = {{Alsing}, Justin and {Peiris}, Hiranya and {Leja}, Joel and {Hahn}, ChangHoon and {Tojeiro}, Rita and others},
        title = "{SPECULATOR: Emulating Stellar Population Synthesis for Fast and Accurate Galaxy Spectra and Photometry}",
      journal = {The Astrophysical Journal Supplement},
         year = 2020,
        month = jul,
       volume = {249},
       number = {1},
          eid = {5},
        pages = {5},
          doi = {10.3847/1538-4365/ab917f},
archivePrefix = {arXiv},
       eprint = {1911.11778},
 primaryClass = {astro-ph.IM},
       adsurl = {https://ui.adsabs.harvard.edu/abs/2020ApJS..249....5A}
}

@ARTICLE{cosmopower,
       author = {{Spurio Mancini}, Alessio and {Piras}, Davide and {Alsing}, Justin and {Joachimi}, Benjamin and {Hobson}, Michael P.},
        title = "{COSMOPOWER: emulating cosmological power spectra for accelerated Bayesian inference from next-generation surveys}",
      journal = {Mon. Notices Royal Astron. Soc.},
         year = 2022,
        month = apr,
       volume = {511},
       number = {2},
        pages = {1771-1788},
          doi = {10.1093/mnras/stac064},
archivePrefix = {arXiv},
       eprint = {2106.03846},
 primaryClass = {astro-ph.CO},
       adsurl = {https://ui.adsabs.harvard.edu/abs/2022MNRAS.511.1771S}
}

@ARTICLE{accelerated-inference-emulator-lsst,
       author = {{Boruah}, Supranta S. and {Eifler}, Tim and {Miranda}, Vivian and {Krishanth}, P.~M. Sai},
        title = "{Accelerating cosmological inference with Gaussian processes and neural networks - an application to LSST Y1 weak lensing and galaxy clustering}",
      journal = {Mon. Notices Royal Astron. Soc.},
         year = 2023,
        month = feb,
       volume = {518},
       number = {4},
        pages = {4818-4831},
          doi = {10.1093/mnras/stac3417},
archivePrefix = {arXiv},
       eprint = {2203.06124},
 primaryClass = {astro-ph.CO},
       adsurl = {https://ui.adsabs.harvard.edu/abs/2023MNRAS.518.4818B}
}

@ARTICLE{euclidemu2,
      author = {{Euclid Collaboration} and {Knabenhans}, M. and {Stadel}, J. and {Potter}, D. and {Dakin}, J. and others},
        title = "{Euclid preparation: IX. EuclidEmulator2 - power spectrum emulation with massive neutrinos and self-consistent dark energy perturbations}",
      journal = {Mon. Not. Royal Astron. Soc.},
         year = 2021,
        month = aug,
       volume = {505},
       number = {2},
        pages = {2840-2869},
          doi = {10.1093/mnras/stab1366},
archivePrefix = {arXiv},
       eprint = {2010.11288},
 primaryClass = {astro-ph.CO},
       adsurl = {https://ui.adsabs.harvard.edu/abs/2021MNRAS.505.2840E}
}

@ARTICLE{lsst-survey-specifications,
      author = {{The LSST Dark Energy Science Collaboration} and {Mandelbaum}, Rachel and {Eifler}, Tim and {Hlo{\v{z}}ek}, Ren{\'e}e and {Collett}, Thomas and others},
        title = "{The LSST Dark Energy Science Collaboration (DESC) Science Requirements Document}",
      journal = {arXiv e-prints},
         year = 2018,
        month = sep,
          eid = {arXiv:1809.01669},
        pages = {arXiv:1809.01669},
          doi = {10.48550/arXiv.1809.01669},
archivePrefix = {arXiv},
       eprint = {1809.01669},
 primaryClass = {astro-ph.CO},
       adsurl = {https://ui.adsabs.harvard.edu/abs/2018arXiv180901669T}
}

@ARTICLE{cosmocov,
       author = {{Fang}, Xiao and {Eifler}, Tim and {Krause}, Elisabeth},
        title = "{2D-FFTLog: efficient computation of real-space covariance matrices for galaxy clustering and weak lensing}",
      journal = {Mon. Notices Royal Astron. Soc.},
         year = 2020,
        month = sep,
       volume = {497},
       number = {3},
        pages = {2699-2714},
          doi = {10.1093/mnras/staa1726},
archivePrefix = {arXiv},
       eprint = {2004.04833},
 primaryClass = {astro-ph.CO},
       adsurl = {https://ui.adsabs.harvard.edu/abs/2020MNRAS.497.2699F}
}

@ARTICLE{bacco,
       author = {{Angulo}, Raul E. and {Zennaro}, Matteo and {Contreras}, Sergio and {Aric{\`o}}, Giovanni and {Pellejero-Iba{\~n}ez}, Marcos and {St{\"u}cker}, Jens},
        title = "{The BACCO simulation project: exploiting the full power of large-scale structure for cosmology}",
      journal = {Mon. Notices Royal Astron. Soc.},
         year = 2021,
        month = nov,
       volume = {507},
       number = {4},
        pages = {5869-5881},
          doi = {10.1093/mnras/stab2018},
archivePrefix = {arXiv},
       eprint = {2004.06245},
 primaryClass = {astro-ph.CO},
       adsurl = {https://ui.adsabs.harvard.edu/abs/2021MNRAS.507.5869A}
}

@ARTICLE{ia-overview,
      author = {{Joachimi}, Benjamin and {Cacciato}, Marcello and {Kitching}, Thomas D. and {Leonard}, Adrienne and {Mandelbaum}, Rachel and others},
        title = "{Galaxy Alignments: An Overview}",
      journal = {Space Science Reviews},
         year = 2015,
        month = nov,
       volume = {193},
       number = {1-4},
        pages = {1-65},
          doi = {10.1007/s11214-015-0177-4},
archivePrefix = {arXiv},
       eprint = {1504.05456},
 primaryClass = {astro-ph.GA},
       adsurl = {https://ui.adsabs.harvard.edu/abs/2015SSRv..193....1J}
}

@ARTICLE{ia-review,
       author = {{Troxel}, M.~A. and {Ishak}, Mustapha},
        title = "{The intrinsic alignment of galaxies and its impact on weak gravitational lensing in an era of precision cosmology}",
      journal = {Physics Reports},
         year = 2015,
        month = feb,
       volume = {558},
        pages = {1-59},
          doi = {10.1016/j.physrep.2014.11.001},
archivePrefix = {arXiv},
       eprint = {1407.6990},
 primaryClass = {astro-ph.CO},
       adsurl = {https://ui.adsabs.harvard.edu/abs/2015PhR...558....1T}
}

@article{Tassev:2013pn,
    author = "Tassev, Svetlin and Zaldarriaga, Matias and Eisenstein, Daniel",
    title = "{Solving Large Scale Structure in Ten Easy Steps with COLA}",
    eprint = "1301.0322",
    archivePrefix = "arXiv",
    primaryClass = "astro-ph.CO",
    doi = "10.1088/1475-7516/2013/06/036",
    journal = "JCAP",
    volume = "06",
    pages = "036",
    year = "2013"
}

@BOOK{PM1988,
       author = {{Hockney}, R.~W. and {Eastwood}, J.~W.},
        title = "{Computer simulation using particles}",
         year = 1988,
       adsurl = {https://ui.adsabs.harvard.edu/abs/1988csup.book.....H}
}

@article{Euclid:2022qde,
    author = "Adamek, J. and others",
    collaboration = "Euclid",
    title = "{Euclid: Modelling massive neutrinos in cosmology \textendash{} a code comparison}",
    eprint = "2211.12457",
    archivePrefix = "arXiv",
    primaryClass = "astro-ph.CO",
    doi = "10.1088/1475-7516/2023/06/035",
    journal = "JCAP",
    volume = "06",
    pages = "035",
    year = "2023"
}

@article{Izard:2015dja,
    author = "Izard, Albert and Crocce, Martin and Fosalba, Pablo",
    title = "{ICE-COLA: Towards fast and accurate synthetic galaxy catalogues optimizing a quasi $N$-body method}",
    eprint = "1509.04685",
    archivePrefix = "arXiv",
    primaryClass = "astro-ph.CO",
    doi = "10.1093/mnras/stw797",
    journal = "Mon. Not. Roy. Astron. Soc.",
    volume = "459",
    number = "3",
    pages = "2327--2341",
    year = "2016"
}

@article{Fiorini:2023fjl,
    author = "Fiorini, Bartolomeo and Koyama, Kazuya and Baker, Tessa",
    title = "{Fast production of cosmological emulators in modified gravity: the matter power spectrum}",
    eprint = "2310.05786",
    archivePrefix = "arXiv",
    primaryClass = "astro-ph.CO",
    doi = "10.1088/1475-7516/2023/12/045",
    journal = "JCAP",
    volume = "12",
    pages = "045",
    year = "2023"
}

@ARTICLE{Brando:2022gvg,
       author = {{Brando}, Guilherme and {Fiorini}, Bartolomeo and {Koyama}, Kazuya and {Winther}, Hans A.},
        title = "{Enabling matter power spectrum emulation in beyond-{\ensuremath{\Lambda}}CDM cosmologies with COLA}",
      journal = {Journal of Cosmology and Astroparticle Physics},
         year = 2022,
        month = sep,
       volume = {2022},
       number = {9},
          eid = {051},
        pages = {051},
          doi = {10.1088/1475-7516/2022/09/051},
archivePrefix = {arXiv},
       eprint = {2203.11120},
 primaryClass = {astro-ph.CO},
       adsurl = {https://ui.adsabs.harvard.edu/abs/2022JCAP...09..051B}
}

@ARTICLE{hicola,
       author = {{Wright}, Bill S. and {Sen Gupta}, Ashim and {Baker}, Tessa and {Valogiannis}, Georgios and {Fiorini}, Bartolomeo and {LSST Dark Energy Science Collaboration}},
        title = "{Hi-COLA: fast, approximate simulations of structure formation in Horndeski gravity}",
      journal = {Journal of Cosmology and Astroparticle Physics},
         year = 2023,
        month = mar,
       volume = {2023},
       number = {3},
          eid = {040},
        pages = {040},
          doi = {10.1088/1475-7516/2023/03/040},
archivePrefix = {arXiv},
       eprint = {2209.01666},
 primaryClass = {astro-ph.CO},
       adsurl = {https://ui.adsabs.harvard.edu/abs/2023JCAP...03..040W}
}

@ARTICLE{pkdgrav3,
       author = {{Potter}, Douglas and {Stadel}, Joachim and {Teyssier}, Romain},
        title = "{PKDGRAV3: beyond trillion particle cosmological simulations for the next era of galaxy surveys}",
      journal = {Computational Astrophysics and Cosmology},
         year = 2017,
        month = may,
       volume = {4},
       number = {1},
          eid = {2},
        pages = {2},
          doi = {10.1186/s40668-017-0021-1},
archivePrefix = {arXiv},
       eprint = {1609.08621},
 primaryClass = {astro-ph.IM},
       adsurl = {https://ui.adsabs.harvard.edu/abs/2017ComAC...4....2P}
}

@article{Angulo_2022,
	doi = {10.1007/s41115-021-00013-z},
  
	url = {https://doi.org/10.1007%2Fs41115-021-00013-z},
  
	year = 2022,
	month = {feb},
  
	publisher = {Springer Science and Business Media {LLC}
},
  
	volume = {8},
  
	number = {1},
  
	author = {Raul E. Angulo and Oliver Hahn},
  
	title = {Large-scale dark matter simulations},
  
	journal = {Living Reviews in Computational Astrophysics}
}

@article{Brando:2020ouk,
    author = "Brando, Guilherme and Koyama, Kazuya and Wands, David",
    title = "{Relativistic Corrections to the Growth of Structure in Modified Gravity}",
    eprint = "2006.11019",
    archivePrefix = "arXiv",
    primaryClass = "astro-ph.CO",
    doi = "10.1088/1475-7516/2021/01/013",
    journal = "JCAP",
    volume = "01",
    pages = "013",
    year = "2021"
}

@article{Tram:2018znz,
    author = "Tram, Thomas and Brandbyge, Jacob and Dakin, Jeppe and Hannestad, Steen",
    title = "{Fully relativistic treatment of light neutrinos in $N$-body simulations}",
    eprint = "1811.00904",
    archivePrefix = "arXiv",
    primaryClass = "astro-ph.CO",
    doi = "10.1088/1475-7516/2019/03/022",
    journal = "JCAP",
    volume = "03",
    pages = "022",
    year = "2019"
}

@article{Fidler:2015npa,
      author = "Fidler, Christian and Rampf, Cornelius and Tram, Thomas and Crittenden, Robert and Koyama, Kazuya and others",
    title = "{General relativistic corrections to $N$-body simulations and the Zel'dovich approximation}",
    eprint = "1505.04756",
    archivePrefix = "arXiv",
    primaryClass = "astro-ph.CO",
    doi = "10.1103/PhysRevD.92.123517",
    journal = "Phys. Rev. D",
    volume = "92",
    number = "12",
    pages = "123517",
    year = "2015"
}

@article{Fidler:2017ebh,
      author = "Fidler, Christian and Tram, Thomas and Rampf, Cornelius and Crittenden, Robert and Koyama, Kazuya and others",
    title = "{Relativistic initial conditions for N-body simulations}",
    eprint = "1702.03221",
    archivePrefix = "arXiv",
    primaryClass = "astro-ph.CO",
    doi = "10.1088/1475-7516/2017/06/043",
    journal = "JCAP",
    volume = "06",
    pages = "043",
    year = "2017"
}

@ARTICLE{cola_mock_galaxy_catalogs,
       author = {{Ding}, Jiacheng and {Li}, Shaohong and {Zheng}, Yi and {Luo}, Xiaolin and {Zhang}, Le and {Li}, Xiao-Dong},
        title = "{Fast generation of mock galaxy catalogues with COLA}",
      journal = {arXiv e-prints},
         year = 2023,
        month = nov,
          eid = {arXiv:2311.00981},
        pages = {arXiv:2311.00981},
          doi = {10.48550/arXiv.2311.00981},
archivePrefix = {arXiv},
       eprint = {2311.00981},
 primaryClass = {astro-ph.CO},
       adsurl = {https://ui.adsabs.harvard.edu/abs/2023arXiv231100981D}
}

@article{DeRose:2018xdj,
      author = "DeRose, Joseph and Wechsler, Risa H. and Tinker, Jeremy L. and Becker, Matthew R. and Mao, Yao-Yuan and others",
    title = "{The Aemulus Project I: Numerical Simulations for Precision Cosmology}",
    eprint = "1804.05865",
    archivePrefix = "arXiv",
    primaryClass = "astro-ph.CO",
    doi = "10.3847/1538-4357/ab1085",
    journal = "Astrophys. J.",
    volume = "875",
    number = "1",
    pages = "69",
    year = "2019"
}

@ARTICLE{gadget4,
       author = {{Springel}, Volker and {Pakmor}, R{\"u}diger and {Zier}, Oliver and {Reinecke}, Martin},
        title = "{Simulating cosmic structure formation with the GADGET-4 code}",
      journal = {\mnras},
         year = 2021,
        month = sep,
       volume = {506},
       number = {2},
        pages = {2871-2949},
          doi = {10.1093/mnras/stab1855},
archivePrefix = {arXiv},
       eprint = {2010.03567},
 primaryClass = {astro-ph.IM},
       adsurl = {https://ui.adsabs.harvard.edu/abs/2021MNRAS.506.2871S}
}

@article{McClintock:2018uyf,
      author = "McClintock, Thomas and Rozo, Eduardo and Becker, Matthew R. and DeRose, Joseph and Mao, Yau-Yuan and others",
    title = "{The Aemulus Project II: Emulating the Halo Mass Function}",
    eprint = "1804.05866",
    archivePrefix = "arXiv",
    primaryClass = "astro-ph.CO",
    doi = "10.3847/1538-4357/aaf568",
    journal = "Astrophys. J.",
    volume = "872",
    number = "1",
    pages = "53",
    year = "2019"
}

@article{Zhai:2018plk,
      author = "Zhai, Zhongxu and Tinker, Jeremy L. and Becker, Matthew R. and DeRose, Joseph and Mao, Yao-Yuan and others",
    title = "{The Aemulus Project III: Emulation of the Galaxy Correlation Function}",
    eprint = "1804.05867",
    archivePrefix = "arXiv",
    primaryClass = "astro-ph.CO",
    doi = "10.3847/1538-4357/ab0d7b",
    journal = "Astrophys. J.",
    volume = "874",
    number = "1",
    pages = "95",
    year = "2019"
}

@article{Takahashi:2012em,
    author = "Takahashi, Ryuichi and Sato, Masanori and Nishimichi, Takahiro and Taruya, Atsushi and Oguri, Masamune",
    title = "{Revising the Halofit Model for the Nonlinear Matter Power Spectrum}",
    eprint = "1208.2701",
    archivePrefix = "arXiv",
    primaryClass = "astro-ph.CO",
    doi = "10.1088/0004-637X/761/2/152",
    journal = "Astrophys. J.",
    volume = "761",
    pages = "152",
    year = "2012"
}

@article{Mead:2015yca,
    author = "Mead, Alexander and Peacock, John and Heymans, Catherine and Joudaki, Shahab and Heavens, Alan",
    title = "{An accurate halo model for fitting non-linear cosmological power spectra and baryonic feedback models}",
    eprint = "1505.07833",
    archivePrefix = "arXiv",
    primaryClass = "astro-ph.CO",
    doi = "10.1093/mnras/stv2036",
    journal = "Mon. Not. Roy. Astron. Soc.",
    volume = "454",
    number = "2",
    pages = "1958--1975",
    year = "2015"
}

@article{Mead:2016zqy,
      author = "Mead, Alexander and Heymans, Catherine and Lombriser, Lucas and Peacock, John and Steele, Olivia and others",
    title = "{Accurate halo-model matter power spectra with dark energy, massive neutrinos and modified gravitational forces}",
    eprint = "1602.02154",
    archivePrefix = "arXiv",
    primaryClass = "astro-ph.CO",
    doi = "10.1093/mnras/stw681",
    journal = "Mon. Not. Roy. Astron. Soc.",
    volume = "459",
    number = "2",
    pages = "1468--1488",
    year = "2016"
}

@ARTICLE{Mead:2020vgs,
       author = {{Mead}, A.~J. and {Brieden}, S. and {Tr{\"o}ster}, T. and {Heymans}, C.},
        title = "{HMCODE-2020: improved modelling of non-linear cosmological power spectra with baryonic feedback}",
      journal = {Monthly Notices of the Royal Astronomical Society},
         year = 2021,
        month = mar,
       volume = {502},
       number = {1},
        pages = {1401-1422},
          doi = {10.1093/mnras/stab082},
archivePrefix = {arXiv},
       eprint = {2009.01858},
 primaryClass = {astro-ph.CO},
       adsurl = {https://ui.adsabs.harvard.edu/abs/2021MNRAS.502.1401M}
}

@ARTICLE{baryonification,
       author = {{Schneider}, Aurel and {Kova{\v{c}}}, Michael and {Bucko}, Jozef and {Nicola}, Andrina and {Reischke}, Robert and {Giri}, Sambit K. and {Teyssier}, Romain and {Tr{\"o}ster}, Tilman and {Refregier}, Alexandre and {Schaller}, Matthieu and {Schaye}, Joop},
        title = "{Baryonification: An alternative to hydrodynamical simulations for cosmological studies}",
      journal = {arXiv e-prints},
         year = 2025,
        month = jul,
          eid = {arXiv:2507.07892},
        pages = {arXiv:2507.07892},
          doi = {10.48550/arXiv.2507.07892},
archivePrefix = {arXiv},
       eprint = {2507.07892},
 primaryClass = {astro-ph.CO},
       adsurl = {https://ui.adsabs.harvard.edu/abs/2025arXiv250707892S}
}

@ARTICLE{cola_baryons,
       author = {{Dai}, Biwei and {Feng}, Yu and {Seljak}, Uro{\v{s}}},
        title = "{A gradient based method for modeling baryons and matter in halos of fast simulations}",
      journal = {\jcap},
         year = 2018,
        month = nov,
       volume = {2018},
       number = {11},
          eid = {009},
        pages = {009},
          doi = {10.1088/1475-7516/2018/11/009},
archivePrefix = {arXiv},
       eprint = {1804.00671},
 primaryClass = {astro-ph.CO},
       adsurl = {https://ui.adsabs.harvard.edu/abs/2018JCAP...11..009D}
}

@ARTICLE{estimating_pk_sims,
       author = {{Colombi}, St{\'e}phane and {Jaffe}, Andrew and {Novikov}, Dmitri and {Pichon}, Christophe},
        title = "{Accurate estimators of power spectra in N-body simulations}",
      journal = {\mnras},
         year = 2009,
        month = feb,
       volume = {393},
       number = {2},
        pages = {511-526},
          doi = {10.1111/j.1365-2966.2008.14176.x},
archivePrefix = {arXiv},
       eprint = {0811.0313},
 primaryClass = {astro-ph},
       adsurl = {https://ui.adsabs.harvard.edu/abs/2009MNRAS.393..511C}
}

@ARTICLE{McEwen:2016fjn,
       author = {{McEwen}, Joseph E. and {Fang}, Xiao and {Hirata}, Christopher M. and {Blazek}, Jonathan A.},
        title = "{FAST-PT: a novel algorithm to calculate convolution integrals in cosmological perturbation theory}",
      journal = {Journal of Cosmology and Astroparticle Physics},
         year = 2016,
        month = sep,
       volume = {2016},
       number = {9},
          eid = {015},
        pages = {015},
          doi = {10.1088/1475-7516/2016/09/015},
archivePrefix = {arXiv},
       eprint = {1603.04826},
 primaryClass = {astro-ph.CO},
       adsurl = {https://ui.adsabs.harvard.edu/abs/2016JCAP...09..015M}
}

@article{Chudaykin:2020aoj,
    author = "Chudaykin, Anton and Ivanov, Mikhail M. and Philcox, Oliver H. E. and Simonovi\'c, Marko",
    title = "{Nonlinear perturbation theory extension of the Boltzmann code CLASS}",
    eprint = "2004.10607",
    archivePrefix = "arXiv",
    primaryClass = "astro-ph.CO",
    reportNumber = "INR-TH-2020-016, CERN-TH-2020-062",
    doi = "10.1103/PhysRevD.102.063533",
    journal = "Phys. Rev. D",
    volume = "102",
    number = "6",
    pages = "063533",
    year = "2020"
}

@ARTICLE{sym-eft,
       author = {{Farakou}, Despoina and {Skordis}, Constantinos},
        title = "{Sym-EFT: Accelerating Effective Field Theory of Large Scale Structure with Symbolic Regression}",
      journal = {arXiv e-prints},
         year = 2025,
        month = nov,
          eid = {arXiv:2511.05093},
        pages = {arXiv:2511.05093},
          doi = {10.48550/arXiv.2511.05093},
archivePrefix = {arXiv},
       eprint = {2511.05093},
 primaryClass = {astro-ph.CO},
       adsurl = {https://ui.adsabs.harvard.edu/abs/2025arXiv251105093F}
}

@ARTICLE{cobra,
       author = {{Bakx}, Thomas and {Chisari}, Nora Elisa and {Vlah}, Zvonimir},
        title = "{Optimal Factorization of Cosmological Large-Scale Structure Observables}",
      journal = {\prl},
         year = 2025,
        month = may,
       volume = {134},
       number = {19},
          eid = {191002},
        pages = {191002},
          doi = {10.1103/PhysRevLett.134.191002},
archivePrefix = {arXiv},
       eprint = {2407.04660},
 primaryClass = {astro-ph.CO},
       adsurl = {https://ui.adsabs.harvard.edu/abs/2025PhRvL.134s1002B}
}

@ARTICLE{DAmico:2020kxu,
       author = {{D'Amico}, Guido and {Senatore}, Leonardo and {Zhang}, Pierre},
        title = "{Limits on wCDM from the EFTofLSS with the PyBird code}",
      journal = {Journal of Cosmology and Astroparticle Physics},
         year = 2021,
        month = jan,
       volume = {2021},
       number = {1},
          eid = {006},
        pages = {006},
          doi = {10.1088/1475-7516/2021/01/006},
archivePrefix = {arXiv},
       eprint = {2003.07956},
 primaryClass = {astro-ph.CO},
       adsurl = {https://ui.adsabs.harvard.edu/abs/2021JCAP...01..006D}
}

@ARTICLE{des_y3_strategy,
      author = {{Krause}, E. and {Fang}, X. and {Pandey}, S. and {Secco}, L.~F. and {Alves}, O. and others},
        title = "{Dark Energy Survey Year 3 Results: Multi-Probe Modeling Strategy and Validation}",
      journal = {arXiv e-prints},
         year = 2021,
        month = may,
          eid = {arXiv:2105.13548},
        pages = {arXiv:2105.13548},
          doi = {10.48550/arXiv.2105.13548},
archivePrefix = {arXiv},
       eprint = {2105.13548},
 primaryClass = {astro-ph.CO},
       adsurl = {https://ui.adsabs.harvard.edu/abs/2021arXiv210513548K}
}

@article{Schneider:2015yka,
      author = "Schneider, Aurel and Teyssier, Romain and Potter, Doug and Stadel, Joachim and Onions, Julian and others",
    title = "{Matter power spectrum and the challenge of percent accuracy}",
    eprint = "1503.05920",
    archivePrefix = "arXiv",
    primaryClass = "astro-ph.CO",
    doi = "10.1088/1475-7516/2016/04/047",
    journal = "JCAP",
    volume = "04",
    pages = "047",
    year = "2016"
}

@ARTICLE{adam_opt,
       author = {{Kingma}, Diederik P. and {Ba}, Jimmy},
        title = "{Adam: A Method for Stochastic Optimization}",
      journal = {arXiv e-prints},
         year = 2014,
        month = dec,
          eid = {arXiv:1412.6980},
        pages = {arXiv:1412.6980},
          doi = {10.48550/arXiv.1412.6980},
archivePrefix = {arXiv},
       eprint = {1412.6980},
 primaryClass = {cs.LG},
       adsurl = {https://ui.adsabs.harvard.edu/abs/2014arXiv1412.6980K}
}

@ARTICLE{desy1_shear,
      author = {{Troxel}, M.~A. and {MacCrann}, N. and {Zuntz}, J. and {Eifler}, T.~F. and {Krause}, E. and others},
        title = "{Dark Energy Survey Year 1 results: Cosmological constraints from cosmic shear}",
      journal = {Phys. Rev. D},
         year = 2018,
        month = aug,
       volume = {98},
       number = {4},
          eid = {043528},
        pages = {043528},
          doi = {10.1103/PhysRevD.98.043528},
archivePrefix = {arXiv},
       eprint = {1708.01538},
 primaryClass = {astro-ph.CO},
       adsurl = {https://ui.adsabs.harvard.edu/abs/2018PhRvD..98d3528T}
}

@ARTICLE{desy1_3x2,
      author = {{Abbott}, T.~M.~C. and {Abdalla}, F.~B. and {Alarcon}, A. and {Aleksi{\'c}}, J. and {Allam}, S. and others},
        title = "{Dark Energy Survey year 1 results: Cosmological constraints from galaxy clustering and weak lensing}",
      journal = {Phys. Rev. D},
         year = 2018,
        month = aug,
       volume = {98},
       number = {4},
          eid = {043526},
        pages = {043526},
          doi = {10.1103/PhysRevD.98.043526},
archivePrefix = {arXiv},
       eprint = {1708.01530},
 primaryClass = {astro-ph.CO},
       adsurl = {https://ui.adsabs.harvard.edu/abs/2018PhRvD..98d3526A}
}

@ARTICLE{desy3_3x2,
      author = {{Abbott}, T.~M.~C. and {Aguena}, M. and {Alarcon}, A. and {Allam}, S. and {Alves}, O. and others},
        title = "{Dark Energy Survey Year 3 results: Cosmological constraints from galaxy clustering and weak lensing}",
      journal = {Phys. Rev. D},
         year = 2022,
        month = jan,
       volume = {105},
       number = {2},
          eid = {023520},
        pages = {023520},
          doi = {10.1103/PhysRevD.105.023520},
archivePrefix = {arXiv},
       eprint = {2105.13549},
 primaryClass = {astro-ph.CO},
       adsurl = {https://ui.adsabs.harvard.edu/abs/2022PhRvD.105b3520A}
}

@ARTICLE{desy3_shear1,
      author = {{Amon}, A. and {Gruen}, D. and {Troxel}, M.~A. and {MacCrann}, N. and {Dodelson}, S. and others},
        title = "{Dark Energy Survey Year 3 results: Cosmology from cosmic shear and robustness to data calibration}",
      journal = {Phys. Rev. D},
         year = 2022,
        month = jan,
       volume = {105},
       number = {2},
          eid = {023514},
        pages = {023514},
          doi = {10.1103/PhysRevD.105.023514},
archivePrefix = {arXiv},
       eprint = {2105.13543},
 primaryClass = {astro-ph.CO},
       adsurl = {https://ui.adsabs.harvard.edu/abs/2022PhRvD.105b3514A}
}

@ARTICLE{desy3_shear2,
      author = {{Secco}, L.~F. and {Samuroff}, S. and {Krause}, E. and {Jain}, B. and {Blazek}, J. and others},
        title = "{Dark Energy Survey Year 3 results: Cosmology from cosmic shear and robustness to modeling uncertainty}",
      journal = {Phys. Rev. D},
         year = 2022,
        month = jan,
       volume = {105},
       number = {2},
          eid = {023515},
        pages = {023515},
          doi = {10.1103/PhysRevD.105.023515},
archivePrefix = {arXiv},
       eprint = {2105.13544},
 primaryClass = {astro-ph.CO},
       adsurl = {https://ui.adsabs.harvard.edu/abs/2022PhRvD.105b3515S}
}

@ARTICLE{desy3_redmagic,
      author = {{Pandey}, S. and {Krause}, E. and {DeRose}, J. and {MacCrann}, N. and {Jain}, B. and others},
        title = "{Dark Energy Survey year 3 results: Constraints on cosmological parameters and galaxy-bias models from galaxy clustering and galaxy-galaxy lensing using the redMaGiC sample}",
      journal = {Phys. Rev. D},
         year = 2022,
        month = aug,
       volume = {106},
       number = {4},
          eid = {043520},
        pages = {043520},
          doi = {10.1103/PhysRevD.106.043520},
archivePrefix = {arXiv},
       eprint = {2105.13545},
 primaryClass = {astro-ph.CO},
       adsurl = {https://ui.adsabs.harvard.edu/abs/2022PhRvD.106d3520P}
}

@ARTICLE{desy3_maglim,
      author = {{Porredon}, A. and {Crocce}, M. and {Elvin-Poole}, J. and {Cawthon}, R. and {Giannini}, G. and others},
        title = "{Dark Energy Survey Year 3 results: Cosmological constraints from galaxy clustering and galaxy-galaxy lensing using the MAGLIM lens sample}",
      journal = {Phys. Rev. D},
         year = 2022,
        month = nov,
       volume = {106},
       number = {10},
          eid = {103530},
        pages = {103530},
          doi = {10.1103/PhysRevD.106.103530},
archivePrefix = {arXiv},
       eprint = {2105.13546},
 primaryClass = {astro-ph.CO},
       adsurl = {https://ui.adsabs.harvard.edu/abs/2022PhRvD.106j3530P}
}

@ARTICLE{des_y1_extensions,
      author = {{Abbott}, T.~M.~C. and {Abdalla}, F.~B. and {Avila}, S. and {Banerji}, M. and {Baxter}, E. and others},
        title = "{Dark Energy Survey year 1 results: Constraints on extended cosmological models from galaxy clustering and weak lensing}",
      journal = {Physical Review D},
         year = 2019,
        month = jun,
       volume = {99},
       number = {12},
          eid = {123505},
        pages = {123505},
          doi = {10.1103/PhysRevD.99.123505},
archivePrefix = {arXiv},
       eprint = {1810.02499},
 primaryClass = {astro-ph.CO},
       adsurl = {https://ui.adsabs.harvard.edu/abs/2019PhRvD..99l3505A}
}

@ARTICLE{des_y3_extensions,
      author = {{Abbott}, T.~M.~C. and {Aguena}, M. and {Alarcon}, A. and {Alves}, O. and {Amon}, A. and others},
        title = "{Dark Energy Survey Year 3 results: Constraints on extensions to {\ensuremath{\Lambda}} CDM with weak lensing and galaxy clustering}",
      journal = {\prd},
         year = 2023,
        month = apr,
       volume = {107},
       number = {8},
          eid = {083504},
        pages = {083504},
          doi = {10.1103/PhysRevD.107.083504},
archivePrefix = {arXiv},
       eprint = {2207.05766},
 primaryClass = {astro-ph.CO},
       adsurl = {https://ui.adsabs.harvard.edu/abs/2023PhRvD.107h3504A}
}

@ARTICLE{camb,
       author = {{Lewis}, Antony and {Challinor}, Anthony and {Lasenby}, Anthony},
        title = "{Efficient Computation of Cosmic Microwave Background Anisotropies in Closed Friedmann-Robertson-Walker Models}",
      journal = {\apj},
         year = 2000,
        month = aug,
       volume = {538},
       number = {2},
        pages = {473-476},
          doi = {10.1086/309179},
archivePrefix = {arXiv},
       eprint = {astro-ph/9911177},
 primaryClass = {astro-ph},
       adsurl = {https://ui.adsabs.harvard.edu/abs/2000ApJ...538..473L}
}

@ARTICLE{pantheonplus,
       author = {{Scolnic}, Dan and {Brout}, Dillon and {Carr}, Anthony and {Riess}, Adam G. and {Davis}, Tamara M. and {Dwomoh}, Arianna and {Jones}, David O. and {Ali}, Noor and {Charvu}, Pranav and {Chen}, Rebecca and {Peterson}, Erik R. and {Popovic}, Brodie and {Rose}, Benjamin M. and {Wood}, Charlotte M. and {Brown}, Peter J. and {Chambers}, Ken and {Coulter}, David A. and {Dettman}, Kyle G. and {Dimitriadis}, Georgios and {Filippenko}, Alexei V. and {Foley}, Ryan J. and {Jha}, Saurabh W. and {Kilpatrick}, Charles D. and {Kirshner}, Robert P. and {Pan}, Yen-Chen and {Rest}, Armin and {Rojas-Bravo}, Cesar and {Siebert}, Matthew R. and {Stahl}, Benjamin E. and {Zheng}, WeiKang},
        title = "{The Pantheon+ Analysis: The Full Data Set and Light-curve Release}",
      journal = {\apj},
         year = 2022,
        month = oct,
       volume = {938},
       number = {2},
          eid = {113},
        pages = {113},
          doi = {10.3847/1538-4357/ac8b7a},
archivePrefix = {arXiv},
       eprint = {2112.03863},
 primaryClass = {astro-ph.CO},
       adsurl = {https://ui.adsabs.harvard.edu/abs/2022ApJ...938..113S}
}

@ARTICLE{web_halo_model,
       author = {{Brieden}, Samuel and {Beutler}, Florian and {Pellejero-Iba{\~n}ez}, Marcos},
        title = "{Web-Halo Model (WHM): Accurate non-linear matter power spectrum predictions without free parameters}",
      journal = {arXiv e-prints},
         year = 2025,
        month = aug,
          eid = {arXiv:2508.10902},
        pages = {arXiv:2508.10902},
          doi = {10.48550/arXiv.2508.10902},
archivePrefix = {arXiv},
       eprint = {2508.10902},
 primaryClass = {astro-ph.CO},
       adsurl = {https://ui.adsabs.harvard.edu/abs/2025arXiv250810902B}
}

@ARTICLE{casarini,
       author = {{Casarini}, Luciano and {Macci{\`o}}, Andrea V. and {Bonometto}, Silvio A.},
        title = "{Dynamical dark energy simulations: high accuracy power spectra at high redshift}",
      journal = {\jcap},
         year = 2009,
        month = mar,
       volume = {2009},
       number = {3},
          eid = {014},
        pages = {014},
          doi = {10.1088/1475-7516/2009/03/014},
archivePrefix = {arXiv},
       eprint = {0810.0190},
 primaryClass = {astro-ph},
       adsurl = {https://ui.adsabs.harvard.edu/abs/2009JCAP...03..014C}
}

@ARTICLE{class,
       author = {{Lesgourgues}, Julien},
        title = "{The Cosmic Linear Anisotropy Solving System (CLASS) I: Overview}",
      journal = {arXiv e-prints},
         year = 2011,
        month = apr,
          eid = {arXiv:1104.2932},
        pages = {arXiv:1104.2932},
          doi = {10.48550/arXiv.1104.2932},
archivePrefix = {arXiv},
       eprint = {1104.2932},
 primaryClass = {astro-ph.IM},
       adsurl = {https://ui.adsabs.harvard.edu/abs/2011arXiv1104.2932L}
}

@ARTICLE{class2,
       author = {{Blas}, Diego and {Lesgourgues}, Julien and {Tram}, Thomas},
        title = "{The Cosmic Linear Anisotropy Solving System (CLASS). Part II: Approximation schemes}",
      journal = {Journal of Cosmology and Astroparticle Physics},
         year = 2011,
        month = jul,
       volume = {2011},
       number = {7},
          eid = {034},
        pages = {034},
          doi = {10.1088/1475-7516/2011/07/034},
archivePrefix = {arXiv},
       eprint = {1104.2933},
 primaryClass = {astro-ph.CO},
       adsurl = {https://ui.adsabs.harvard.edu/abs/2011JCAP...07..034B}
}

@ARTICLE{desy3_cov,
      author = {{Friedrich}, O. and {Andrade-Oliveira}, F. and {Camacho}, H. and {Alves}, O. and {Rosenfeld}, R. and others},
        title = "{Dark Energy Survey year 3 results: covariance modelling and its impact on parameter estimation and quality of fit}",
      journal = {Mon. Not. of the Royal Astron. Soc.},
         year = 2021,
        month = dec,
       volume = {508},
       number = {3},
        pages = {3125-3165},
          doi = {10.1093/mnras/stab2384},
archivePrefix = {arXiv},
       eprint = {2012.08568},
 primaryClass = {astro-ph.CO},
       adsurl = {https://ui.adsabs.harvard.edu/abs/2021MNRAS.508.3125F}
}

@article{DeRose:2023dmk,
      author = "DeRose, Joseph and Kokron, Nickolas and Banerjee, Arka and Chen, Shi-Fan and White, Martin and others",
    title = "{Aemulus \ensuremath{\nu}: precise predictions for matter and biased tracer power spectra in the presence of neutrinos}",
    eprint = "2303.09762",
    archivePrefix = "arXiv",
    primaryClass = "astro-ph.CO",
    doi = "10.1088/1475-7516/2023/07/054",
    journal = "JCAP",
    volume = "07",
    pages = "054",
    year = "2023"
}

@ARTICLE{desi_dr1,
      author = {{DESI Collaboration} and {Abdul-Karim}, M. and {Adame}, A.~G. and {Aguado}, D. and {Aguilar}, J. and others},
        title = "{Data Release 1 of the Dark Energy Spectroscopic Instrument}",
      journal = {arXiv e-prints},
         year = 2025,
        month = mar,
          eid = {arXiv:2503.14745},
        pages = {arXiv:2503.14745},
          doi = {10.48550/arXiv.2503.14745},
archivePrefix = {arXiv},
       eprint = {2503.14745},
 primaryClass = {astro-ph.CO},
       adsurl = {https://ui.adsabs.harvard.edu/abs/2025arXiv250314745D}
}

@ARTICLE{desi_dr2_results_lya,
      author = {{DESI Collaboration} and {Abdul-Karim}, M. and {Aguilar}, J. and {Ahlen}, S. and {Allende Prieto}, C. and others},
        title = "{DESI DR2 Results I: Baryon Acoustic Oscillations from the Lyman Alpha Forest}",
      journal = {arXiv e-prints},
         year = 2025,
        month = mar,
          eid = {arXiv:2503.14739},
        pages = {arXiv:2503.14739},
          doi = {10.48550/arXiv.2503.14739},
archivePrefix = {arXiv},
       eprint = {2503.14739},
 primaryClass = {astro-ph.CO},
       adsurl = {https://ui.adsabs.harvard.edu/abs/2025arXiv250314739D}
}

@ARTICLE{desi_dr2_bao,
      author = {{DESI Collaboration} and {Abdul-Karim}, M. and {Aguilar}, J. and {Ahlen}, S. and {Alam}, S. and others},
        title = "{DESI DR2 Results II: Measurements of Baryon Acoustic Oscillations and Cosmological Constraints}",
      journal = {arXiv e-prints},
         year = 2025,
        month = mar,
          eid = {arXiv:2503.14738},
        pages = {arXiv:2503.14738},
          doi = {10.48550/arXiv.2503.14738},
archivePrefix = {arXiv},
       eprint = {2503.14738},
 primaryClass = {astro-ph.CO},
       adsurl = {https://ui.adsabs.harvard.edu/abs/2025arXiv250314738D}
}

@ARTICLE{desi_dr2_validation,
      author = {{Andrade}, U. and {Paillas}, E. and {Mena-Fernandez}, J. and {Li}, Q. and {Ross}, A.~J. and others},
        title = "{Validation of the DESI DR2 Measurements of Baryon Acoustic Oscillations from Galaxies and Quasars}",
      journal = {arXiv e-prints},
         year = 2025,
        month = mar,
          eid = {arXiv:2503.14742},
        pages = {arXiv:2503.14742},
          doi = {10.48550/arXiv.2503.14742},
archivePrefix = {arXiv},
       eprint = {2503.14742},
 primaryClass = {astro-ph.CO},
       adsurl = {https://ui.adsabs.harvard.edu/abs/2025arXiv250314742A}
}

@ARTICLE{desi_dr2_validation_lya,
      author = {{Casas}, L. and {Herrera-Alcantar}, H.~K. and {Chaves-Montero}, J. and {Cuceu}, A. and {Font-Ribera}, A. and others},
        title = "{Validation of the DESI DR2 Ly$\alpha$ BAO analysis using synthetic datasets}",
      journal = {arXiv e-prints},
         year = 2025,
        month = mar,
          eid = {arXiv:2503.14741},
        pages = {arXiv:2503.14741},
          doi = {10.48550/arXiv.2503.14741},
archivePrefix = {arXiv},
       eprint = {2503.14741},
 primaryClass = {astro-ph.IM},
       adsurl = {https://ui.adsabs.harvard.edu/abs/2025arXiv250314741C}
}

@ARTICLE{desi_dr2_extensions,
      author = {{Lodha}, K. and {Calderon}, R. and {Matthewson}, W.~L. and {Shafieloo}, A. and {Ishak}, M. and others},
        title = "{Extended Dark Energy analysis using DESI DR2 BAO measurements}",
      journal = {arXiv e-prints},
         year = 2025,
        month = mar,
          eid = {arXiv:2503.14743},
        pages = {arXiv:2503.14743},
          doi = {10.48550/arXiv.2503.14743},
archivePrefix = {arXiv},
       eprint = {2503.14743},
 primaryClass = {astro-ph.CO},
       adsurl = {https://ui.adsabs.harvard.edu/abs/2025arXiv250314743L}
}

@ARTICLE{desi_dr2_neutrinos,
      author = {{Elbers}, W. and {Aviles}, A. and {Noriega}, H.~E. and {Chebat}, D. and {Menegas}, A. and others},
        title = "{Constraints on Neutrino Physics from DESI DR2 BAO and DR1 Full Shape}",
      journal = {arXiv e-prints},
         year = 2025,
        month = mar,
          eid = {arXiv:2503.14744},
        pages = {arXiv:2503.14744},
          doi = {10.48550/arXiv.2503.14744},
archivePrefix = {arXiv},
       eprint = {2503.14744},
 primaryClass = {astro-ph.CO},
       adsurl = {https://ui.adsabs.harvard.edu/abs/2025arXiv250314744E}
}

@ARTICLE{boss_bao,
      author = {{Ross}, Ashley J. and {Beutler}, Florian and {Chuang}, Chia-Hsun and {Pellejero-Ibanez}, Marcos and {Seo}, Hee-Jong and others},
        title = "{The clustering of galaxies in the completed SDSS-III Baryon Oscillation Spectroscopic Survey: observational systematics and baryon acoustic oscillations in the correlation function}",
      journal = {\mnras},
         year = 2017,
        month = jan,
       volume = {464},
       number = {1},
        pages = {1168-1191},
          doi = {10.1093/mnras/stw2372},
archivePrefix = {arXiv},
       eprint = {1607.03145},
 primaryClass = {astro-ph.CO},
       adsurl = {https://ui.adsabs.harvard.edu/abs/2017MNRAS.464.1168R}
}

@ARTICLE{boss_bao_harmonic,
      author = {{Beutler}, Florian and {Seo}, Hee-Jong and {Ross}, Ashley J. and {McDonald}, Patrick and {Saito}, Shun and others},
        title = "{The clustering of galaxies in the completed SDSS-III Baryon Oscillation Spectroscopic Survey: baryon acoustic oscillations in the Fourier space}",
      journal = {\mnras},
         year = 2017,
        month = jan,
       volume = {464},
       number = {3},
        pages = {3409-3430},
          doi = {10.1093/mnras/stw2373},
archivePrefix = {arXiv},
       eprint = {1607.03149},
 primaryClass = {astro-ph.CO},
       adsurl = {https://ui.adsabs.harvard.edu/abs/2017MNRAS.464.3409B}
}

@ARTICLE{boss_rsd,
      author = {{Satpathy}, Siddharth and {Alam}, Shadab and {Ho}, Shirley and {White}, Martin and {Bahcall}, Neta A. and others},
        title = "{The clustering of galaxies in the completed SDSS-III Baryon Oscillation Spectroscopic Survey: on the measurement of growth rate using galaxy correlation functions}",
      journal = {\mnras},
         year = 2017,
        month = aug,
       volume = {469},
       number = {2},
        pages = {1369-1382},
          doi = {10.1093/mnras/stx883},
archivePrefix = {arXiv},
       eprint = {1607.03148},
 primaryClass = {astro-ph.CO},
       adsurl = {https://ui.adsabs.harvard.edu/abs/2017MNRAS.469.1369S}
}

@ARTICLE{boss_rsd_harmonic,
      author = {{Beutler}, Florian and {Seo}, Hee-Jong and {Saito}, Shun and {Chuang}, Chia-Hsun and {Cuesta}, Antonio J. and others},
        title = "{The clustering of galaxies in the completed SDSS-III Baryon Oscillation Spectroscopic Survey: anisotropic galaxy clustering in Fourier space}",
      journal = {\mnras},
         year = 2017,
        month = apr,
       volume = {466},
       number = {2},
        pages = {2242-2260},
          doi = {10.1093/mnras/stw3298},
archivePrefix = {arXiv},
       eprint = {1607.03150},
 primaryClass = {astro-ph.CO},
       adsurl = {https://ui.adsabs.harvard.edu/abs/2017MNRAS.466.2242B}
}

@ARTICLE{boss_results,
      author = {{Alam}, Shadab and {Ata}, Metin and {Bailey}, Stephen and {Beutler}, Florian and {Bizyaev}, Dmitry and others},
        title = "{The clustering of galaxies in the completed SDSS-III Baryon Oscillation Spectroscopic Survey: cosmological analysis of the DR12 galaxy sample}",
      journal = {\mnras},
         year = 2017,
        month = sep,
       volume = {470},
       number = {3},
        pages = {2617-2652},
          doi = {10.1093/mnras/stx721},
archivePrefix = {arXiv},
       eprint = {1607.03155},
 primaryClass = {astro-ph.CO},
       adsurl = {https://ui.adsabs.harvard.edu/abs/2017MNRAS.470.2617A}
}

@ARTICLE{eboss_qso_selection,
      author = {{Myers}, Adam D. and {Palanque-Delabrouille}, Nathalie and {Prakash}, Abhishek and {P{\^a}ris}, Isabelle and {Yeche}, Christophe and others},
        title = "{The SDSS-IV Extended Baryon Oscillation Spectroscopic Survey: Quasar Target Selection}",
      journal = {\apjs},
         year = 2015,
        month = dec,
       volume = {221},
       number = {2},
          eid = {27},
        pages = {27},
          doi = {10.1088/0067-0049/221/2/27},
archivePrefix = {arXiv},
       eprint = {1508.04472},
 primaryClass = {astro-ph.CO},
       adsurl = {https://ui.adsabs.harvard.edu/abs/2015ApJS..221...27M}
}

@ARTICLE{eboss_overview,
      author = {{Dawson}, Kyle S. and {Kneib}, Jean-Paul and {Percival}, Will J. and {Alam}, Shadab and {Albareti}, Franco D. and others},
        title = "{The SDSS-IV Extended Baryon Oscillation Spectroscopic Survey: Overview and Early Data}",
      journal = {\aj},
         year = 2016,
        month = feb,
       volume = {151},
       number = {2},
          eid = {44},
        pages = {44},
          doi = {10.3847/0004-6256/151/2/44},
archivePrefix = {arXiv},
       eprint = {1508.04473},
 primaryClass = {astro-ph.CO},
       adsurl = {https://ui.adsabs.harvard.edu/abs/2016AJ....151...44D}
}

@ARTICLE{eboss_forecast,
      author = {{Zhao}, Gong-Bo and {Wang}, Yuting and {Ross}, Ashley J. and {Shandera}, Sarah and {Percival}, Will J. and others},
        title = "{The extended Baryon Oscillation Spectroscopic Survey: a cosmological forecast}",
      journal = {\mnras},
         year = 2016,
        month = apr,
       volume = {457},
       number = {3},
        pages = {2377-2390},
          doi = {10.1093/mnras/stw135},
archivePrefix = {arXiv},
       eprint = {1510.08216},
 primaryClass = {astro-ph.CO},
       adsurl = {https://ui.adsabs.harvard.edu/abs/2016MNRAS.457.2377Z}
}

@ARTICLE{eboss_qso_clustering,
      author = {{Rodr{\'\i}guez-Torres}, Sergio A. and {Comparat}, Johan and {Prada}, Francisco and {Yepes}, Gustavo and {Burtin}, Etienne and others},
        title = "{Clustering of quasars in the first year of the SDSS-IV eBOSS survey: interpretation and halo occupation distribution}",
      journal = {\mnras},
         year = 2017,
        month = jun,
       volume = {468},
       number = {1},
        pages = {728-740},
          doi = {10.1093/mnras/stx454},
archivePrefix = {arXiv},
       eprint = {1612.06918},
 primaryClass = {astro-ph.CO},
       adsurl = {https://ui.adsabs.harvard.edu/abs/2017MNRAS.468..728R}
}

@ARTICLE{eboss_photoz_des,
      author = {{Jouvel}, S. and {Delubac}, T. and {Comparat}, J. and {Camacho}, H. and {Carnero}, A. and others},
        title = "{Photometric redshifts and clustering of emission line galaxies selected jointly by DES and eBOSS}",
      journal = {\mnras},
         year = 2017,
        month = aug,
       volume = {469},
       number = {3},
        pages = {2771-2790},
          doi = {10.1093/mnras/stx163},
       adsurl = {https://ui.adsabs.harvard.edu/abs/2017MNRAS.469.2771J}
}

@ARTICLE{eboss_lrg,
      author = {{Zhai}, Zhongxu and {Tinker}, Jeremy L. and {Hahn}, ChangHoon and {Seo}, Hee-Jong and {Blanton}, Michael R. and others},
        title = "{The Clustering of Luminous Red Galaxies at z {\ensuremath{\sim}} 0.7 from EBOSS and BOSS Data}",
      journal = {\apj},
         year = 2017,
        month = oct,
       volume = {848},
       number = {2},
          eid = {76},
        pages = {76},
          doi = {10.3847/1538-4357/aa8eee},
archivePrefix = {arXiv},
       eprint = {1607.05383},
 primaryClass = {astro-ph.CO},
       adsurl = {https://ui.adsabs.harvard.edu/abs/2017ApJ...848...76Z}
}

@ARTICLE{eboss_qso_bao,
      author = {{Ata}, Metin and {Baumgarten}, Falk and {Bautista}, Julian and {Beutler}, Florian and {Bizyaev}, Dmitry and others},
        title = "{The clustering of the SDSS-IV extended Baryon Oscillation Spectroscopic Survey DR14 quasar sample: first measurement of baryon acoustic oscillations between redshift 0.8 and 2.2}",
      journal = {\mnras},
         year = 2018,
        month = feb,
       volume = {473},
       number = {4},
        pages = {4773-4794},
          doi = {10.1093/mnras/stx2630},
archivePrefix = {arXiv},
       eprint = {1705.06373},
 primaryClass = {astro-ph.CO},
       adsurl = {https://ui.adsabs.harvard.edu/abs/2018MNRAS.473.4773A}
}

@ARTICLE{eboss_rsd_qso,
      author = {{Gil-Mar{\'\i}n}, H{\'e}ctor and {Guy}, Julien and {Zarrouk}, Pauline and {Burtin}, Etienne and {Chuang}, Chia-Hsun and others},
        title = "{The clustering of the SDSS-IV extended Baryon Oscillation Spectroscopic Survey DR14 quasar sample: structure growth rate measurement from the anisotropic quasar power spectrum in the redshift range 0.8 < z < 2.2}",
      journal = {\mnras},
         year = 2018,
        month = jun,
       volume = {477},
       number = {2},
        pages = {1604-1638},
          doi = {10.1093/mnras/sty453},
archivePrefix = {arXiv},
       eprint = {1801.02689},
 primaryClass = {astro-ph.CO},
       adsurl = {https://ui.adsabs.harvard.edu/abs/2018MNRAS.477.1604G}
}

@ARTICLE{eboss_rsd_qso_2,
      author = {{Ruggeri}, Rossana and {Percival}, Will J. and {Gil-Mar{\'\i}n}, H{\'e}ctor and {Beutler}, Florian and {Mueller}, Eva-Maria and others},
        title = "{The clustering of the SDSS-IV extended Baryon Oscillation Spectroscopic Survey DR14 quasar sample: measuring the evolution of the growth rate using redshift-space distortions between redshift 0.8 and 2.2}",
      journal = {\mnras},
         year = 2019,
        month = mar,
       volume = {483},
       number = {3},
        pages = {3878-3887},
          doi = {10.1093/mnras/sty3395},
archivePrefix = {arXiv},
       eprint = {1801.02891},
 primaryClass = {astro-ph.CO},
       adsurl = {https://ui.adsabs.harvard.edu/abs/2019MNRAS.483.3878R}
}

@ARTICLE{kids-legacy-calibration,
      author = {{Li}, Shun-Sheng and {Kuijken}, Konrad and {Hoekstra}, Henk and {Miller}, Lance and {Heymans}, Catherine and others},
        title = "{KiDS-Legacy calibration: Unifying shear and redshift calibration with the SKiLLS multi-band image simulations}",
      journal = {\aap},
         year = 2023,
        month = feb,
       volume = {670},
          eid = {A100},
        pages = {A100},
          doi = {10.1051/0004-6361/202245210},
archivePrefix = {arXiv},
       eprint = {2210.07163},
 primaryClass = {astro-ph.CO},
       adsurl = {https://ui.adsabs.harvard.edu/abs/2023A&A...670A.100L}
}

@ARTICLE{kids-legacy,
      author = {{Wright}, Angus H. and {St{\"o}lzner}, Benjamin and {Asgari}, Marika and {Bilicki}, Maciej and {Giblin}, Benjamin and others},
        title = "{KiDS-Legacy: Cosmological constraints from cosmic shear with the complete Kilo-Degree Survey}",
      journal = {arXiv e-prints},
         year = 2025,
        month = mar,
          eid = {arXiv:2503.19441},
        pages = {arXiv:2503.19441},
          doi = {10.48550/arXiv.2503.19441},
archivePrefix = {arXiv},
       eprint = {2503.19441},
 primaryClass = {astro-ph.CO},
       adsurl = {https://ui.adsabs.harvard.edu/abs/2025arXiv250319441W}
}

@ARTICLE{kids1000_density_split,
      author = {{Burger}, Pierre A. and {Friedrich}, Oliver and {Harnois-D{\'e}raps}, Joachim and {Schneider}, Peter and {Asgari}, Marika and others},
        title = "{KiDS-1000 cosmology: Constraints from density split statistics}",
      journal = {\aap},
         year = 2023,
        month = jan,
       volume = {669},
          eid = {A69},
        pages = {A69},
          doi = {10.1051/0004-6361/202244673},
archivePrefix = {arXiv},
       eprint = {2208.02171},
 primaryClass = {astro-ph.CO},
       adsurl = {https://ui.adsabs.harvard.edu/abs/2023A&A...669A..69B}
}

@ARTICLE{kids1000_x_cmb,
      author = {{Yao}, Ji and {Shan}, Huanyuan and {Zhang}, Pengjie and {Liu}, Xiangkun and {Heymans}, Catherine and others},
        title = "{KiDS-1000: Cross-correlation with Planck cosmic microwave background lensing and intrinsic alignment removal with self-calibration}",
      journal = {\aap},
         year = 2023,
        month = may,
       volume = {673},
          eid = {A111},
        pages = {A111},
          doi = {10.1051/0004-6361/202346020},
archivePrefix = {arXiv},
       eprint = {2301.13437},
 primaryClass = {astro-ph.CO},
       adsurl = {https://ui.adsabs.harvard.edu/abs/2023A&A...673A.111Y}
}

@ARTICLE{kids1000_halo_constraints,
      author = {{Dvornik}, Andrej and {Heymans}, Catherine and {Asgari}, Marika and {Mahony}, Constance and {Joachimi}, Benjamin and others},
        title = "{KiDS-1000: Combined halo-model cosmology constraints from galaxy abundance, galaxy clustering, and galaxy-galaxy lensing}",
      journal = {\aap},
         year = 2023,
        month = jul,
       volume = {675},
          eid = {A189},
        pages = {A189},
          doi = {10.1051/0004-6361/202245158},
archivePrefix = {arXiv},
       eprint = {2210.03110},
 primaryClass = {astro-ph.CO},
       adsurl = {https://ui.adsabs.harvard.edu/abs/2023A&A...675A.189D}
}

@ARTICLE{kids_euclid_pseudocl,
      author = {{Loureiro}, A. and {Whittaker}, L. and {Spurio Mancini}, A. and {Joachimi}, B. and {Cuceu}, A. and others},
        title = "{KiDS and Euclid: Cosmological implications of a pseudo angular power spectrum analysis of KiDS-1000 cosmic shear tomography}",
      journal = {\aap},
         year = 2022,
        month = sep,
       volume = {665},
          eid = {A56},
        pages = {A56},
          doi = {10.1051/0004-6361/202142481},
archivePrefix = {arXiv},
       eprint = {2110.06947},
 primaryClass = {astro-ph.CO},
       adsurl = {https://ui.adsabs.harvard.edu/abs/2022A&A...665A..56L}
}

@ARTICLE{kids1000_enhance_cal,
      author = {{van den Busch}, J.~L. and {Wright}, A.~H. and {Hildebrandt}, H. and {Bilicki}, M. and {Asgari}, M. and others},
        title = "{KiDS-1000: Cosmic shear with enhanced redshift calibration}",
      journal = {\aap},
         year = 2022,
        month = aug,
       volume = {664},
          eid = {A170},
        pages = {A170},
          doi = {10.1051/0004-6361/202142083},
archivePrefix = {arXiv},
       eprint = {2204.02396},
 primaryClass = {astro-ph.CO},
       adsurl = {https://ui.adsabs.harvard.edu/abs/2022A&A...664A.170V}
}

@ARTICLE{kids1000_ia_lrg,
      author = {{Fortuna}, Maria Cristina and {Hoekstra}, Henk and {Johnston}, Harry and {Vakili}, Mohammadjavad and {Kannawadi}, Arun and others},
        title = "{KiDS-1000: Constraints on the intrinsic alignment of luminous red galaxies}",
      journal = {\aap},
         year = 2021,
        month = oct,
       volume = {654},
          eid = {A76},
        pages = {A76},
          doi = {10.1051/0004-6361/202140706},
archivePrefix = {arXiv},
       eprint = {2109.02556},
 primaryClass = {astro-ph.CO},
       adsurl = {https://ui.adsabs.harvard.edu/abs/2021A&A...654A..76F}
}

@ARTICLE{kids1000_bgs,
      author = {{Bilicki}, M. and {Dvornik}, A. and {Hoekstra}, H. and {Wright}, A.~H. and {Chisari}, N.~E. and others},
        title = "{Bright galaxy sample in the Kilo-Degree Survey Data Release 4. Selection, photometric redshifts, and physical properties}",
      journal = {\aap},
         year = 2021,
        month = sep,
       volume = {653},
          eid = {A82},
        pages = {A82},
          doi = {10.1051/0004-6361/202140352},
archivePrefix = {arXiv},
       eprint = {2101.06010},
 primaryClass = {astro-ph.GA},
       adsurl = {https://ui.adsabs.harvard.edu/abs/2021A&A...653A..82B}
}

@ARTICLE{kids1000_extensions,
      author = {{Tr{\"o}ster}, Tilman and {Asgari}, Marika and {Blake}, Chris and {Cataneo}, Matteo and {Heymans}, Catherine and others},
        title = "{KiDS-1000 Cosmology: Constraints beyond flat {\ensuremath{\Lambda}}CDM}",
      journal = {\aap},
         year = 2021,
        month = may,
       volume = {649},
          eid = {A88},
        pages = {A88},
          doi = {10.1051/0004-6361/202039805},
archivePrefix = {arXiv},
       eprint = {2010.16416},
 primaryClass = {astro-ph.CO},
       adsurl = {https://ui.adsabs.harvard.edu/abs/2021A&A...649A..88T}
}

@ARTICLE{kids1000_qso,
      author = {{Nakoneczny}, S.~J. and {Bilicki}, M. and {Pollo}, A. and {Asgari}, M. and {Dvornik}, A. and others},
        title = "{Photometric selection and redshifts for quasars in the Kilo-Degree Survey Data Release 4}",
      journal = {\aap},
         year = 2021,
        month = may,
       volume = {649},
          eid = {A81},
        pages = {A81},
          doi = {10.1051/0004-6361/202039684},
archivePrefix = {arXiv},
       eprint = {2010.13857},
 primaryClass = {astro-ph.CO},
       adsurl = {https://ui.adsabs.harvard.edu/abs/2021A&A...649A..81N}
}

@ARTICLE{kids1000_photoz_cal,
      author = {{Hildebrandt}, H. and {van den Busch}, J.~L. and {Wright}, A.~H. and {Blake}, C. and {Joachimi}, B. and others},
        title = "{KiDS-1000 catalogue: Redshift distributions and their calibration}",
      journal = {\aap},
         year = 2021,
        month = mar,
       volume = {647},
          eid = {A124},
        pages = {A124},
          doi = {10.1051/0004-6361/202039018},
archivePrefix = {arXiv},
       eprint = {2007.15635},
 primaryClass = {astro-ph.CO},
       adsurl = {https://ui.adsabs.harvard.edu/abs/2021A&A...647A.124H}
}

@ARTICLE{kids-1000,
      author = {{Heymans}, Catherine and {Tr{\"o}ster}, Tilman and {Asgari}, Marika and {Blake}, Chris and {Hildebrandt}, Hendrik and others},
        title = "{KiDS-1000 Cosmology: Multi-probe weak gravitational lensing and spectroscopic galaxy clustering constraints}",
      journal = {Astronomy and Astrophysics},
         year = 2021,
        month = feb,
       volume = {646},
          eid = {A140},
        pages = {A140},
          doi = {10.1051/0004-6361/202039063},
archivePrefix = {arXiv},
       eprint = {2007.15632},
 primaryClass = {astro-ph.CO},
       adsurl = {https://ui.adsabs.harvard.edu/abs/2021A&A...646A.140H}
}

@ARTICLE{wigglez_bao,
      author = {{Blake}, Chris and {Kazin}, Eyal A. and {Beutler}, Florian and {Davis}, Tamara M. and {Parkinson}, David and others},
        title = "{The WiggleZ Dark Energy Survey: mapping the distance-redshift relation with baryon acoustic oscillations}",
      journal = {\mnras},
         year = 2011,
        month = dec,
       volume = {418},
       number = {3},
        pages = {1707-1724},
          doi = {10.1111/j.1365-2966.2011.19592.x},
archivePrefix = {arXiv},
       eprint = {1108.2635},
 primaryClass = {astro-ph.CO},
       adsurl = {https://ui.adsabs.harvard.edu/abs/2011MNRAS.418.1707B}
}

@ARTICLE{wigglez_rsd,
      author = {{Blake}, Chris and {Brough}, Sarah and {Colless}, Matthew and {Contreras}, Carlos and {Couch}, Warrick and others},
        title = "{The WiggleZ Dark Energy Survey: the growth rate of cosmic structure since redshift z=0.9}",
      journal = {\mnras},
         year = 2011,
        month = aug,
       volume = {415},
       number = {3},
        pages = {2876-2891},
          doi = {10.1111/j.1365-2966.2011.18903.x},
archivePrefix = {arXiv},
       eprint = {1104.2948},
 primaryClass = {astro-ph.CO},
       adsurl = {https://ui.adsabs.harvard.edu/abs/2011MNRAS.415.2876B}
}

@ARTICLE{wigglez_bao_2,
      author = {{Kazin}, Eyal A. and {Koda}, Jun and {Blake}, Chris and {Padmanabhan}, Nikhil and {Brough}, Sarah and others},
        title = "{The WiggleZ Dark Energy Survey: improved distance measurements to z = 1 with reconstruction of the baryonic acoustic feature}",
      journal = {\mnras},
         year = 2014,
        month = jul,
       volume = {441},
       number = {4},
        pages = {3524-3542},
          doi = {10.1093/mnras/stu778},
archivePrefix = {arXiv},
       eprint = {1401.0358},
 primaryClass = {astro-ph.CO},
       adsurl = {https://ui.adsabs.harvard.edu/abs/2014MNRAS.441.3524K}
}

\:
\appendix

\section{Simulations and Emulator Accuracy for Higher Redshifts}
\label{app:errors_higher_z}
In this Appendix, we show relative errors from the simulations and the emulator for higher redshifts, ensuring that our discussion in Section~\ref{sec:Methodology} still holds. The results are presented in Figure~\ref{fig:errors_higher_z}. We find that for $z \leq 3$, the $z=0$ results still hold in that at $k = 1 \,h/\mathrm{Mpc}$, 90\% of cosmologies have emulation errors within 2\% on $\tilde{B}(k,z)$, and that 50\% of cosmologies are contained well within 1\%. COLA emulators trained at different redshifts demonstrate no notable change in the fidelity of predictions across the range $0 \leq z \leq 3$ (right panels), even though the highest 10\% of raw emulation errors (left panels) and COLA boost errors (middle panels) grow with increasing $z$.

\begin{figure*}
    \centering  
    \includegraphics[width=0.99\linewidth]{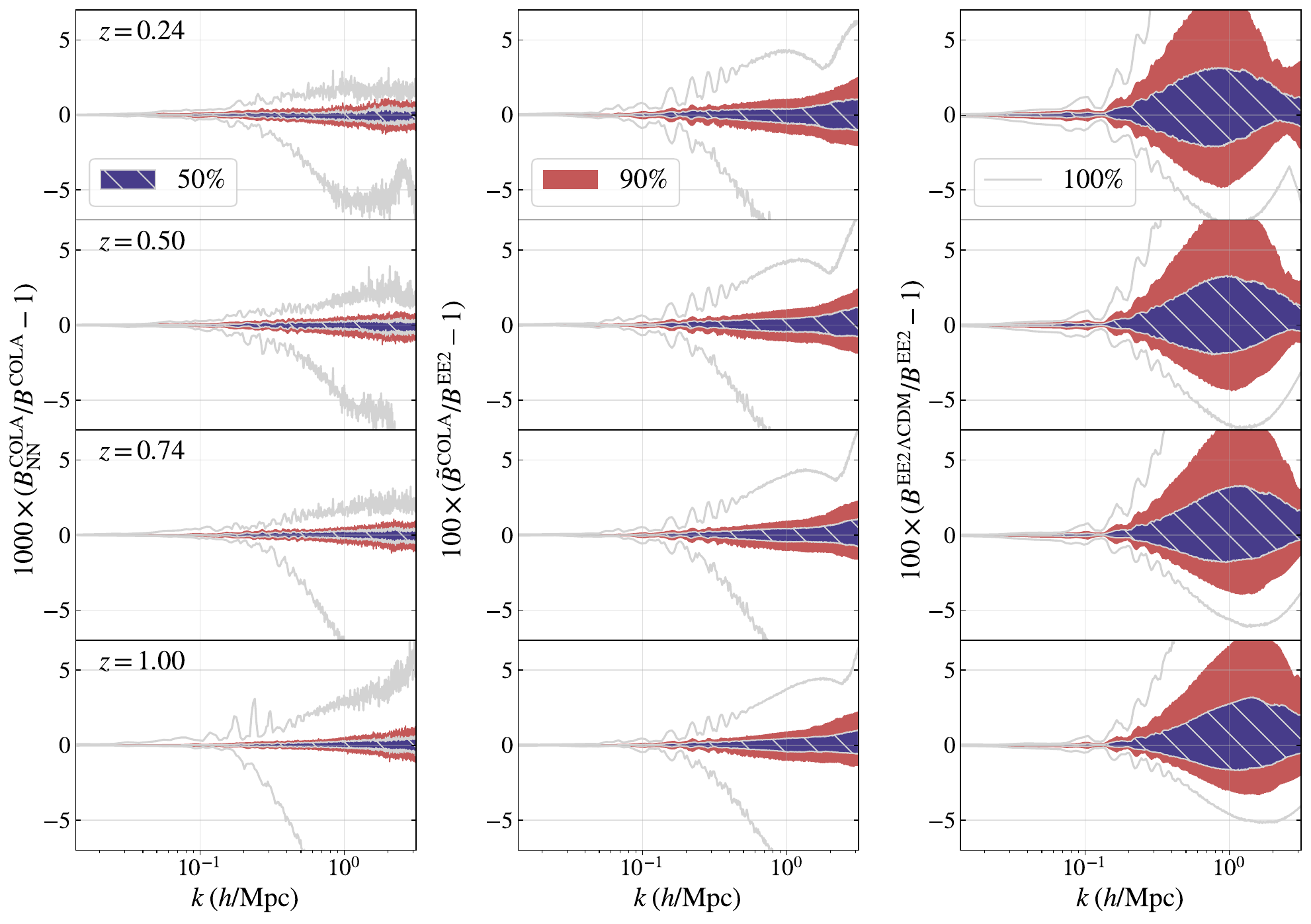}
    \caption{\textbf{Left panel:} Relative errors between the COLA boosts $B^\mathrm{COLA}$ predicted by the emulator versus those obtained from the test set simulations. \textbf{Middle panel:} Relative errors between $\tilde{B}^\mathrm{COLA}$ (see Equation~\ref{eq:inf_refs}) versus the boosts from \euclidemutwo. \textbf{Right panel:} Relative errors between $B^{\mathrm{EE2} \; \Lambda \mathrm{CDM}}$ and \eetwo. Each row denotes a different redshift. Colors in all panels denote the percentile of cosmologies around the mean: blue contours enclose $50\%$ of cosmologies, red contours enclose $90\%$ of cosmologies, and the outer gray lines enclose $100\%$ of cosmologies.}
    \label{fig:errors_higher_z}
\end{figure*}

\end{document}